\documentclass[times,twocolumn,final]{elsarticle}

\usepackage{url}
\usepackage{hyperref}
\hypersetup{hidelinks}

\usepackage{cag}
\usepackage{framed,multirow}

\usepackage{latexsym}

\newcommand{\eg}{\emph{e.g.}\xspace}

\newcommand{\myparagraph}[1]{\smallskip\noindent\textbf{#1}\xspace}
\newcommand{\systemname}{\textsc{UrbanGazeVis}\xspace}
\newcommand{\dataname}{\textsc{UrbanGaze}\xspace}

\usepackage[english]{babel}
\usepackage{hyperref}
\addto\extrasenglish{%
}

\usepackage{textcomp}
\usepackage[T1]{fontenc}
\usepackage{xspace}
\usepackage{amssymb}
\usepackage{enumitem}
\usepackage{threeparttable}
\usepackage{amsmath}
\usepackage{subfig}
\usepackage{svg}
\usepackage[table,dvipsnames]{xcolor}
\usepackage{booktabs}
\usepackage{colortbl}
\usepackage[normalem]{ulem}
\usepackage{tabularx}
\usepackage{hyperref}
\usepackage[normalem]{ulem}

\definecolor{newcolor}{rgb}{.8,.349,.1}
\definecolor{lightgray}{gray}{0.95}

\usepackage[switch,pagewise]{lineno} 

\journal{Computers \& Graphics}

\begin{document}

\verso{Preprint Submitted for review}

\begin{frontmatter}

\title{UrbanGazeVis: A Visualization System for Analyzing Eye-Tracking Data on Urban Safety Perception}%

\author[1]{Andres \snm{De La Puente}\corref{cor1}}

 \emailauthor{andres.puente@fgv.br}{Andres \snm{De La Puente}\corref{cor1}}

 \author[1]{Luis \snm{Sante}\fnref{fn1}}
    
 \author[1]{Felipe \snm{Moreno-Vera}\fnref{fn1}}
 \author[1]{Mauro \snm{Diaz}\fnref{fn1}}
 \author[1]{Jorge \snm{Poco}\fnref{fn1}}
 \address[1]{Fundação Getulio Vargas. Praia de Botafogo, 190 - Botafogo, Rio de Janeiro - RJ, 22250-145, Brazil}

\received{\today}

\begin{abstract}
Perceived safety in streetscapes depends on where people look, yet how gaze relates to visual cues of urban disorder remains poorly understood. Prior work treats safety as an image-level label, offering little insight into how attention to specific elements (e.g, buildings, greenery, people, signs of decay) shapes these judgments. We present a head-mounted eye-tracking study in which 30 participants viewed and rated the safety of 150 street-view images from Rio de Janeiro using a HoloLens 2 headset. Gaze traces were mapped onto semantic segments and disorder cues (e.g., damaged walls, graffiti, overhead cables), yielding a multimodal dataset linking gaze dynamics, scene semantics, and safety scores. To analyze it, we introduce \systemname, an interactive visual analytics system with image- and participant-centric views that connects the spatial, temporal, and semantic dimensions of gaze to perceived safety, supporting comparisons between safe and unsafe scenes, inspection of divergent ratings for similar images, and region-of-interest analysis via glyph-based summaries. Statistical models show that sustained attention to physical disorder is associated with lower perceived safety, while the visual analysis reveals context-specific effects often masked by global aggregation. Together, these analyses offer actionable insights for urban design and planning.
\end{abstract}

\begin{keyword}
\KWD Urban Safety Perception, Visual attention, Eye-tracking, Urban Physical Disorder, Street-level imagery
\end{keyword}

\end{frontmatter}

\newcommand{\figWorkflow}{
\begin{figure}[t!]
    \centering
    \includegraphics[width=\columnwidth]{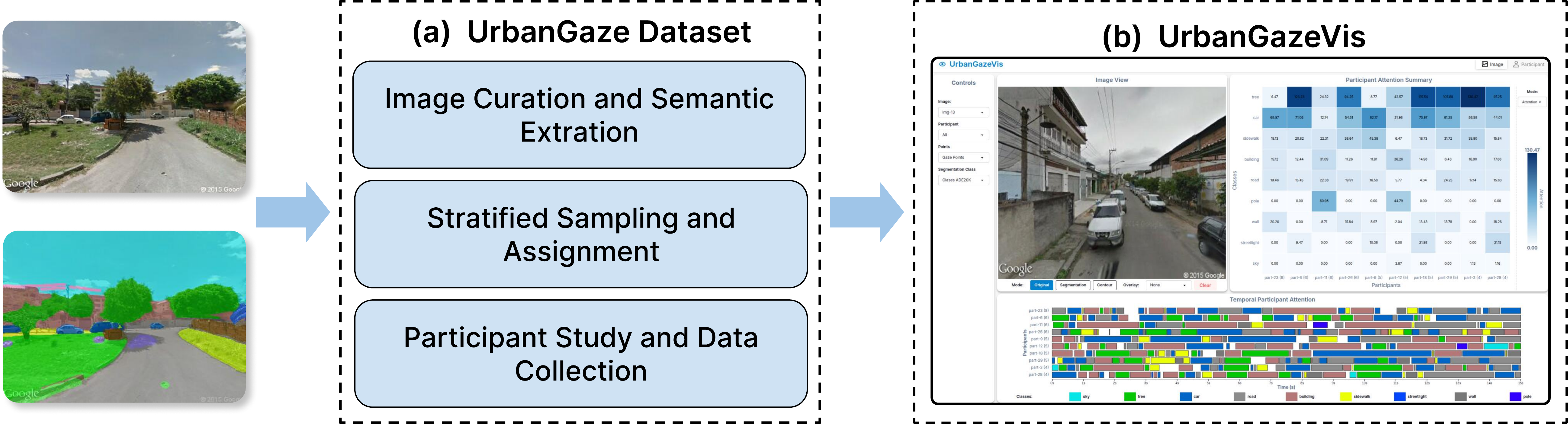}
    \caption{\systemname workflow. 
    \textbf{(a)} Dataset construction: street-view scenes are semantically processed, sampled, and used in the participant study to collect gaze data and safety ratings. 
    \textbf{(b)} System analysis: the resulting multimodal data are explored interactively to analyze spatial and temporal patterns of visual attention and perceived safety.}
    \vspace{-0.4cm}
    \label{fig:workflow}
\end{figure}
}

\newcommand{\figImageFamiliarization}{
\begin{figure}[t!]
        \centering
        \scalebox{0.95}{
        \subfloat[Safe sample.]{\includegraphics[width=0.48\columnwidth]{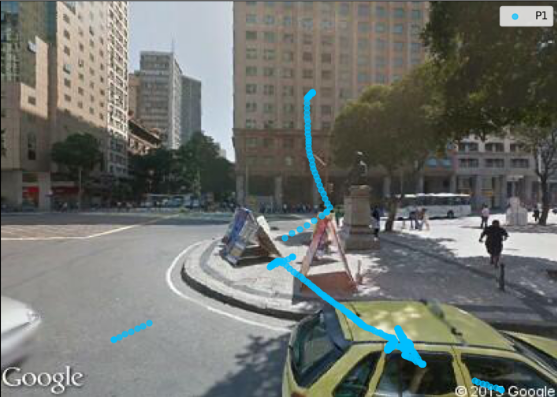}}
        \hfill 
        \subfloat[Unsafe sample.]{\includegraphics[width=0.48\columnwidth]{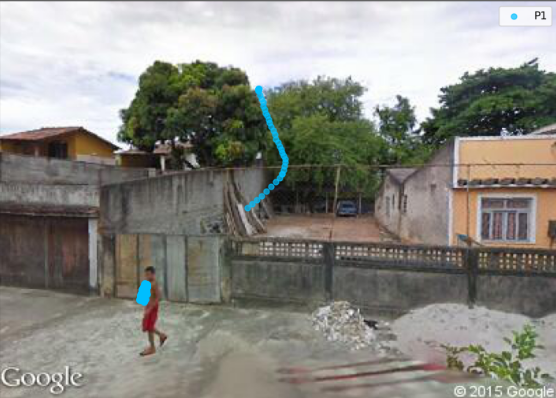}}
        }
        \caption{Example images used during the familiarization phase: \textbf{(a)} a scene perceived as safe and \textbf{(b)} a scene perceived as unsafe. These examples help participants understand the safety-rating task and the 1--10 Likert scale.}
        \vspace{-0.4cm}
        \label{fig:image_familiarization}
    \end{figure} 
}

\newcommand{\figPerceptionAssessment}{
\begin{figure*}[t!]
    \centering
    \includegraphics[width=0.95\textwidth]{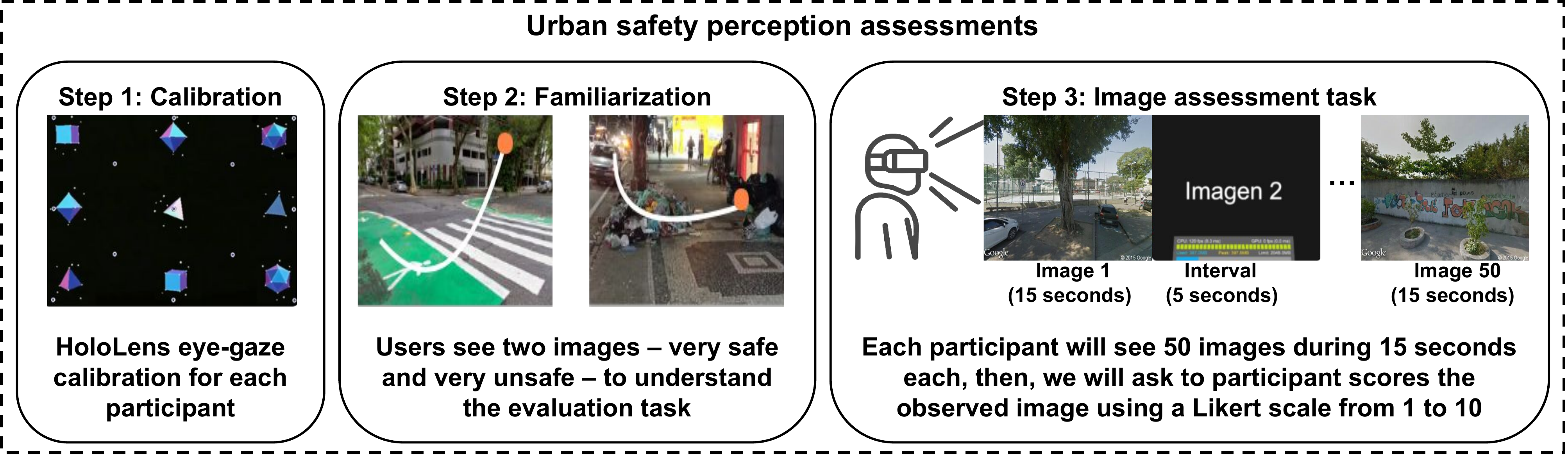}
    \caption{Urban safety perception protocol. 
    Step~1: gaze calibration with the HoloLens~2 headset. 
    Step~2: task familiarization using one clearly safe and one clearly unsafe example image. 
    Step~3: image assessment, in which each participant views 50 assigned images for 15\,s, followed by a 5\,s rating phase in which perceived safety is reported on a 1--10 Likert scale.}
    \vspace{-0.4cm}
    \label{fig:perception_assessment}
\end{figure*}
}

\newcommand{\figScoresPP}{
\begin{figure}[t!]
    \centering
    \includegraphics[width=0.7\columnwidth]{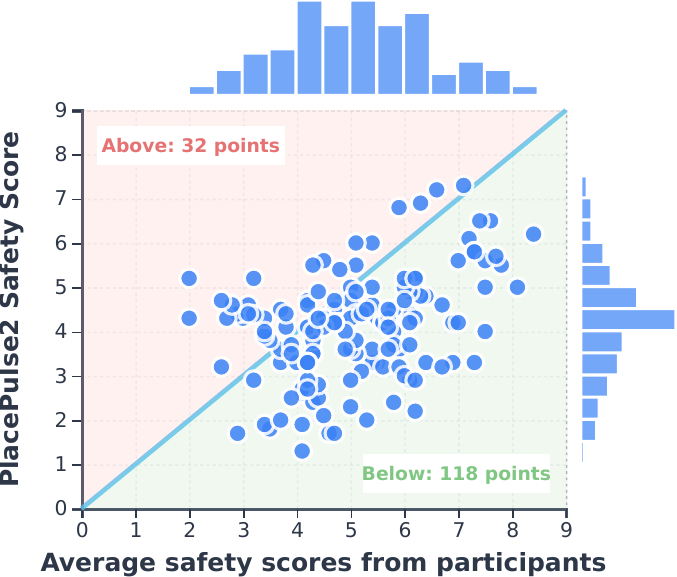} 
    \caption{Comparison between Place Pulse~2.0 safety scores and our mixed-reality ratings. Each point represents one of the 150 images, plotting its Place Pulse score (y-axis) against the mean score from our participants (x-axis). Points below the diagonal correspond to scenes rated as safer by our participants than by the global Place Pulse crowd; points above the diagonal indicate the opposite.} 
    \vspace{-0.4cm}
    \label{fig:scatterplot} 
\end{figure}
}

\newcommand{\figImageAnalyzer}{
\begin{figure*}[ht!]
    \centering
    \includegraphics[width=0.9\textwidth]{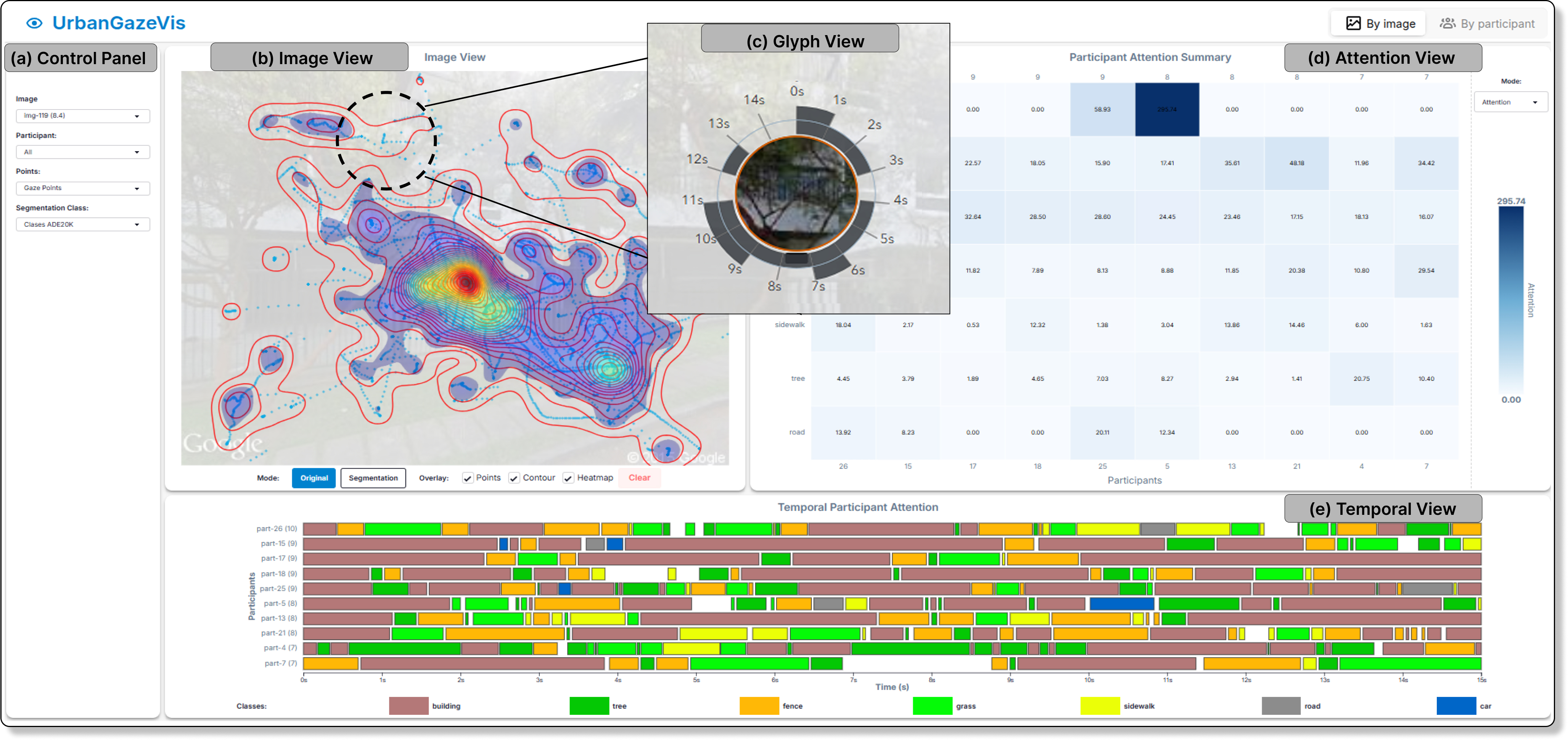}
    \caption{By-Image component of \systemname. 
\textbf{(a)} Control panel for selecting the image, participant filters, point type (gaze or fixation), and semantic representation. 
\textbf{(b)} Image View showing the selected streetscape with optional overlays for gaze points, density contours, and heatmap, and access to the Glyph View for selected regions. 
\textbf{(c)} Glyph View, triggered from a circular region of interest in the Image View, summarizing visual exploration within the selected bounds over the 15\,s trial: stacked radial bars encode participant counts per 1\,s segment.
\textbf{(d)} Attention View (\textit{Participant Attention Summary}) summarizing \textit{Total Attention} or \textit{normalized Attention Intensity (AttIn)} by participant (columns) and semantic class (rows). 
\textbf{(e)} Temporal View (\textit{Temporal Participant Attention}) using a scarf-timeline chart to show, for each participant, which semantic class is fixated at each moment during the 15\,s viewing period. Participant traceability is maintained interactively via tooltips and single-user filtering.}
    \vspace{-0.4cm}
    \label{fig:image_analyzer_component}
\end{figure*}
}

\newcommand{\figParticipantAnalyzer}{
\begin{figure*}[t!]
    \centering
    \includegraphics[width=0.95\textwidth]{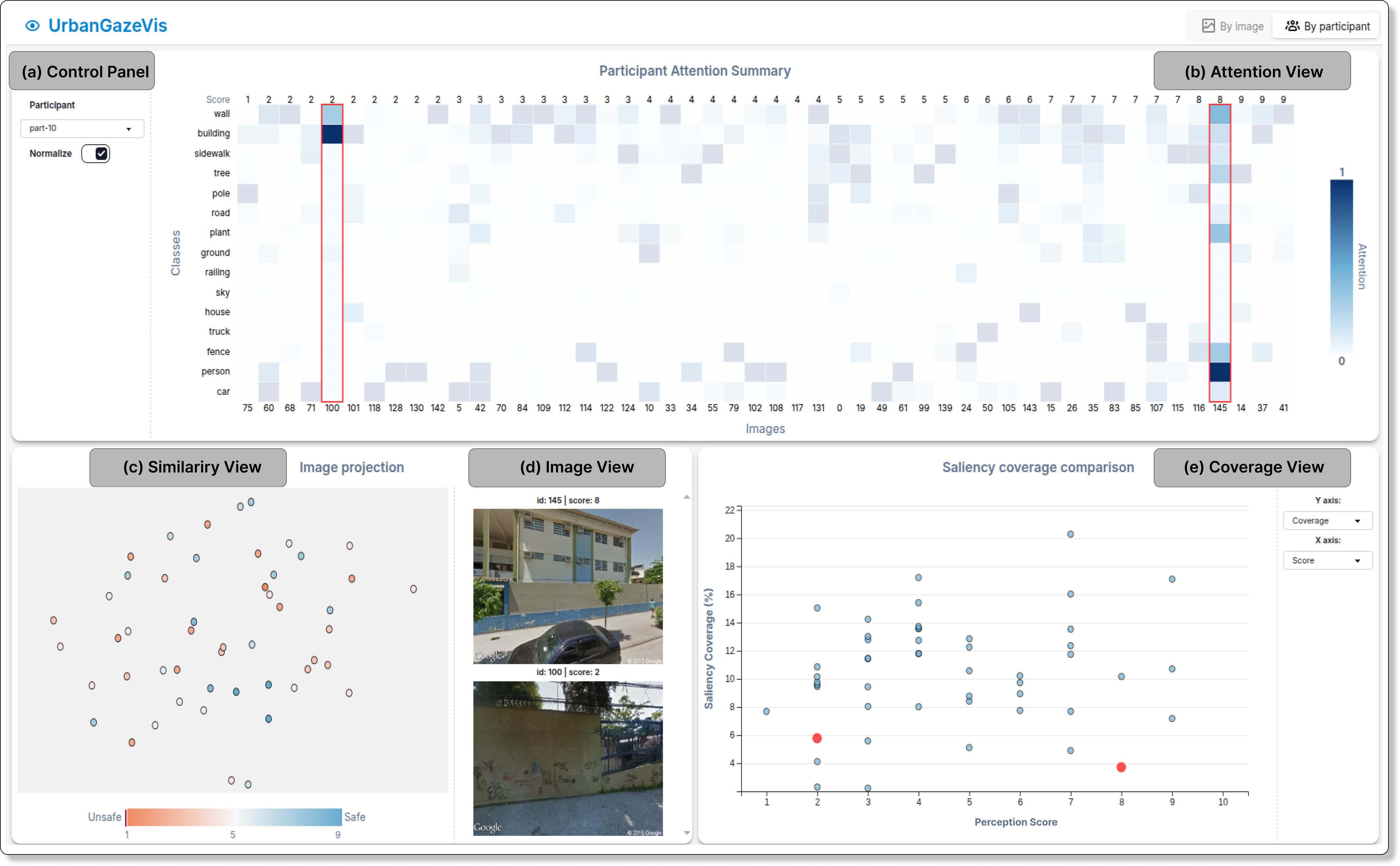}
    \caption{By-Participant component of \systemname{}. 
    \textbf{(a)} Control panel for selecting a participant, choosing the semantic representation, and enabling normalization in the attention heatmap. 
    \textbf{(b)} Attention View showing, for the selected participant, \textit{normalized Attention Intensity (AttIn)} by image (columns, ordered by the participant's safety score) and semantic class (rows). 
    \textbf{(c)} Similarity View (\textit{Image Projection}), where each point represents a Places365 scene embedding projected via t-SNE, colored by the participant's safety score (red = low, blue = high).
    \textbf{(d)} Image View showing the images selected from the projection. 
    \textbf{(e)} Coverage View (\textit{Saliency Coverage Comparison}) relating \textit{Saliency Coverage} or entropy to safety scores or image order.}
    \vspace{-0.4cm}
    \label{fig:participant_analyzer_component}
\end{figure*}
}

\newcommand{\figScenarioA}{
\begin{figure*}[ht!]
    \vspace{0.4cm}
    \centering
    \includegraphics[width=0.9\textwidth]{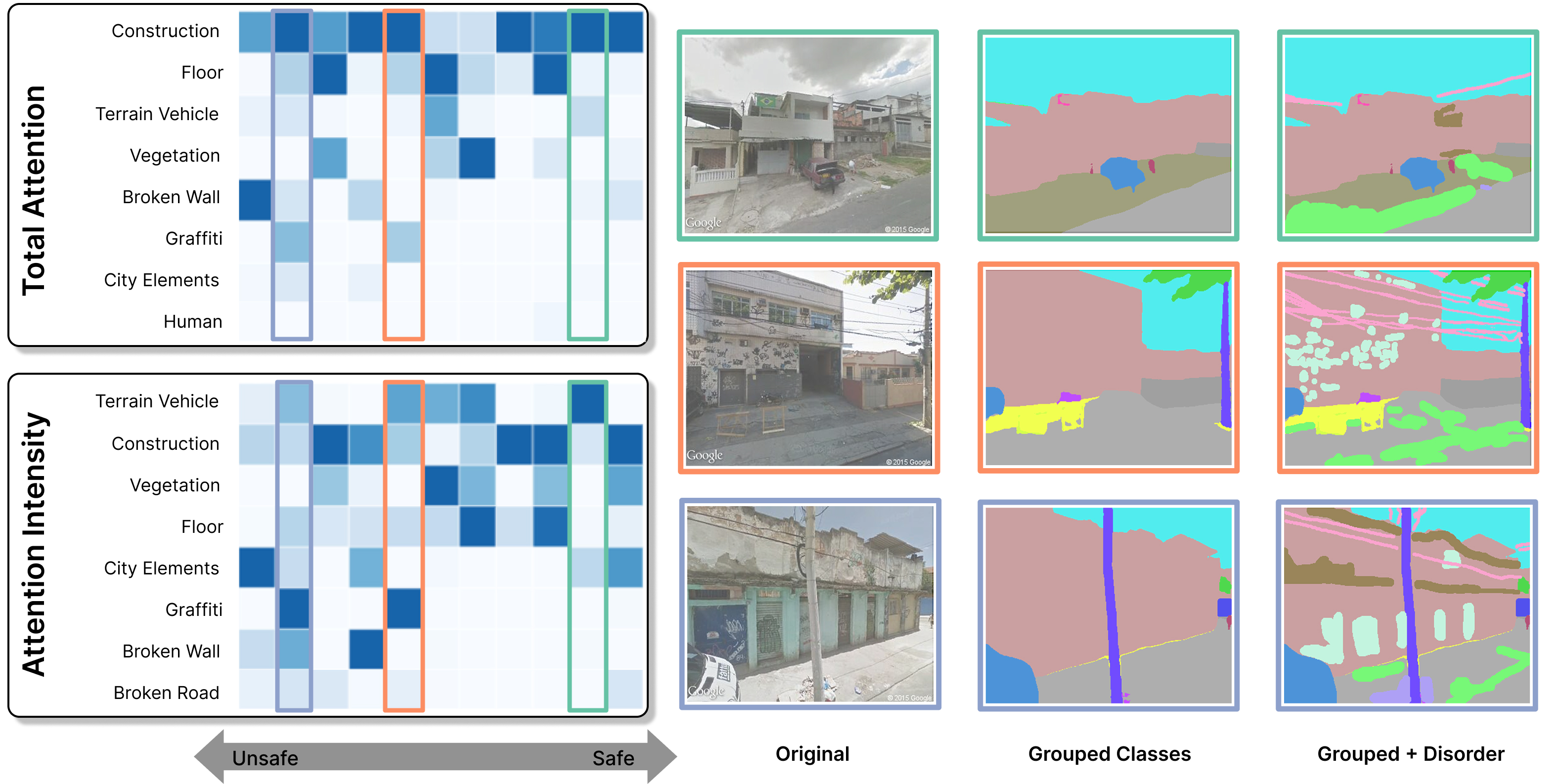} 
\caption{Usage scenario 1: semantic comparison in the By-Participant view for Participant 29. \textbf{(Left)} The Attention View shows two stacked matrices---\textbf{Total Attention} (top) and \textbf{normalized Attention Intensity (AttIn)} (bottom)---with semantic classes as rows (independently ranked by magnitude in each matrix) and the participant's assessed images as columns, ordered left to right from unsafe to safe. Three representative scenes are highlighted with colored outlines (\textcolor[HTML]{8DA0CB}{blue}, \textcolor[HTML]{FC8D62}{orange}, \textcolor[HTML]{66C2A5}{green}). 
 \textbf{(Right)} The three highlighted scenes (Image IDs 60, 38, and 109) are shown one per row---each as the original street-view stimulus \textbf{(left)}, the Grouped classes \textbf{(center)}, and the Grouped + Disorder classes \textbf{(right)}---with row outline colors matching the highlighted columns on the left. Contrasting Total Attention with AttIn enables comparison between attention to general urban structure and to specific disorder cues.} 
    \vspace{0.4cm}
    \label{fig:scenario_1} 
\end{figure*}
}

\newcommand{\figScenarioB}{
\begin{figure*}[ht!]
    \centering
    \includegraphics[width=0.85\textwidth]{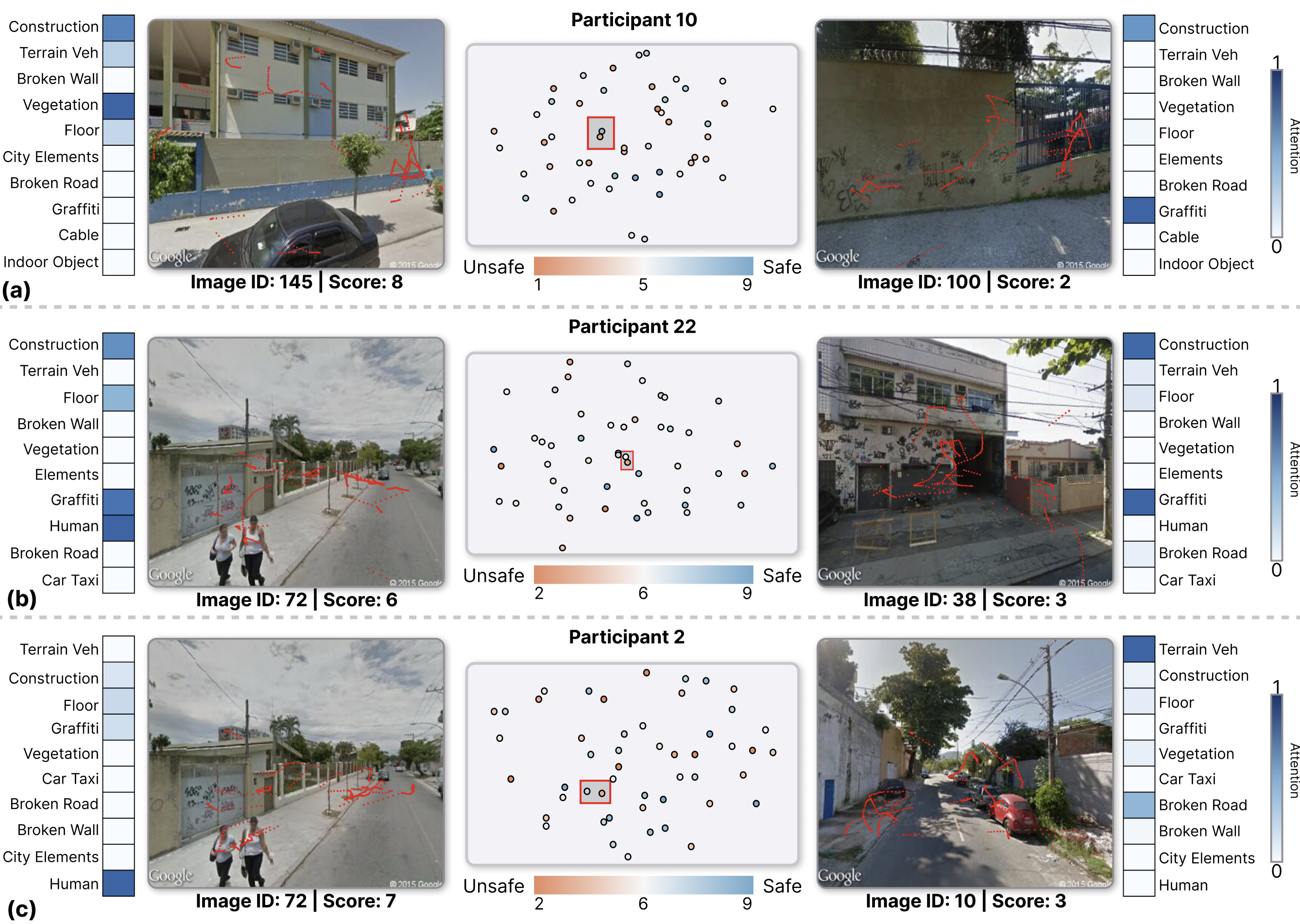} 
\caption{Usage scenario 2: divergent safety ratings for visually similar scenes in the By-Participant view. Using the Similarity View, we identify image pairs (red boxes) that are close in the projection space but received markedly different scores. The side panels summarize the participant's \textbf{Attention Intensity (AttIn)} across visual categories such as vegetation, graffiti, and vehicles. Heatmap intensity is normalized according to the AttIn metric defined in \autoref{eq:attin_formula}.}
    \vspace{-0.4cm}
    \label{fig:usage_case_2} 
\end{figure*}
}

\newcommand{\figScenarioC}{
\begin{figure*}[ht!]
    \centering
    \includegraphics[width=0.93\textwidth]{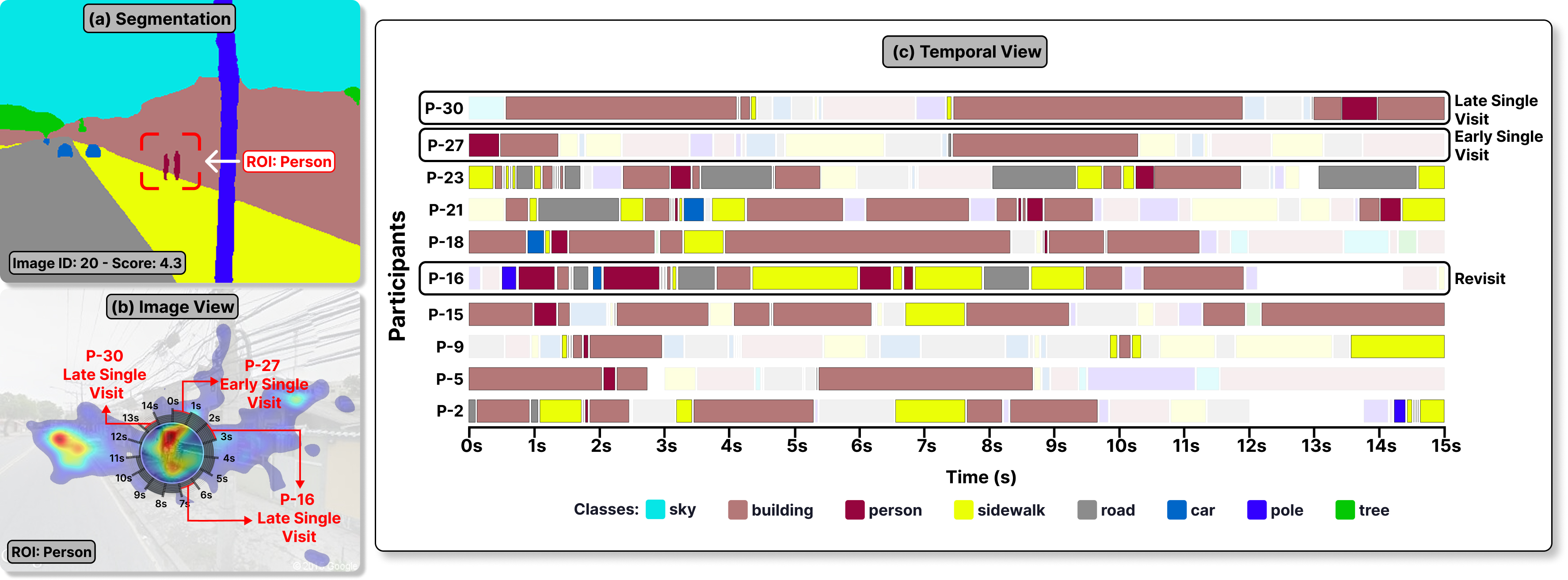}
    \caption{Usage scenario 3: Temporal divergence in exploring the ``person'' class (Image ID 20, safety score: 4.3). \textbf{(a) Semantic segmentation} of Image ID 20. \textbf{(b) Image View} heatmap reveals a high fixation density (visual hotspot) over the pedestrian region, prompting the placement of the ROI Glyph. \textbf{(c) Temporal View} details how individual participants interact with this specific area over the 15-second trial: Participant 27 makes an early single visit, Participant 30 makes a late single visit, and Participant 16 exhibits a recurrent attention (last single visit or revisit) pattern. These distinct spatiotemporal strategies are effectively disentangled, revealing behavioral nuances that the aggregated spatial heatmap alone would conceal.}
    \vspace{-0.4cm}
    \label{fig:scenario_3} \label{fig:spatiotemporal_revisits}
\end{figure*}
}

\newcommand{\figIceChart}{
\begin{figure}[t!]
    \centering
    \includegraphics[width=\columnwidth]{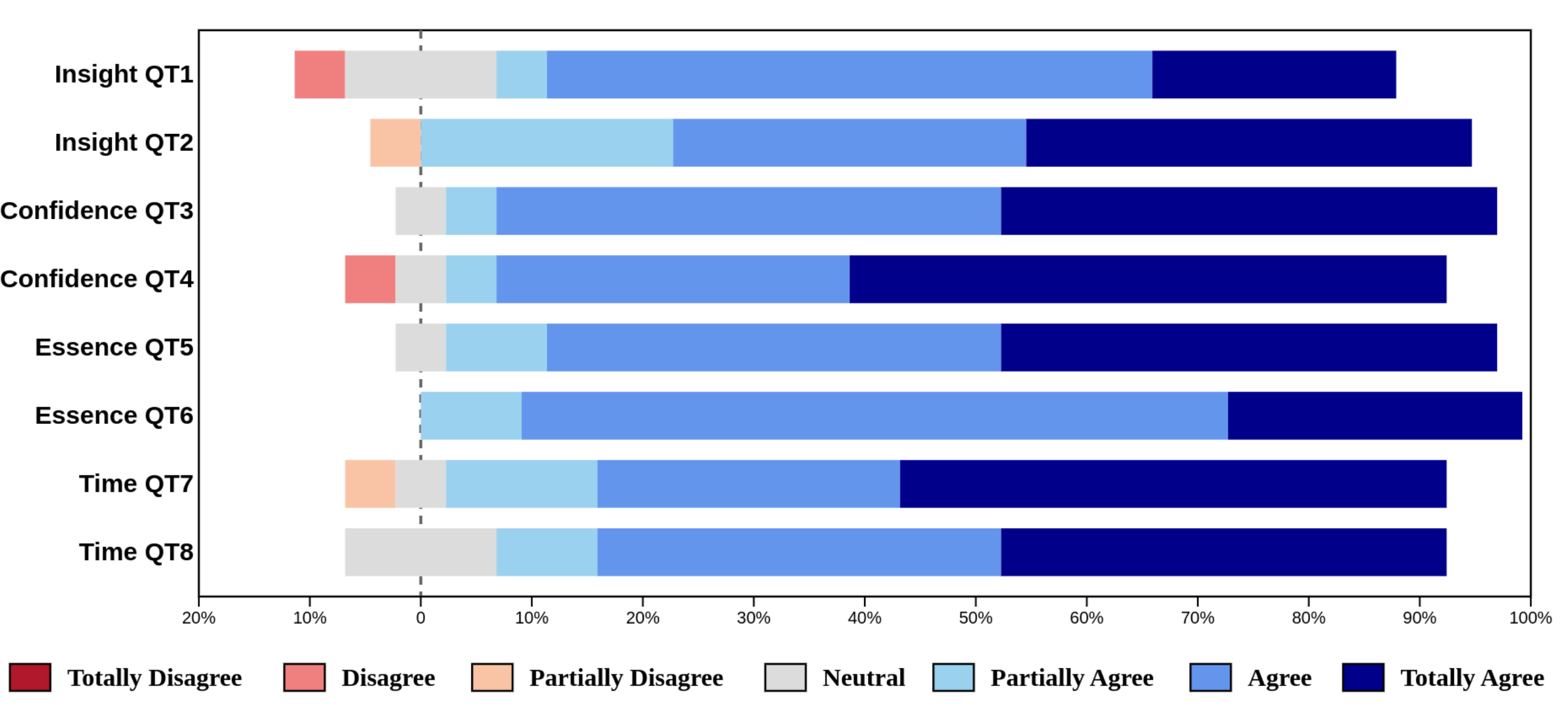} 
    \caption{Distribution of participant responses for questions QT1--QT8, grouped by ICE-T-inspired dimensions (Insight, Confidence, Essence, and Time). Bar length indicates the percentage of participants selecting each Likert level on a 7-point scale from ``Totally Disagree'' to ``Totally Agree''.}
    \vspace{-0.4cm}
    \label{fig:icet_chart} 
\end{figure}
}

\newcommand{\figLmmBaseFull}{
\begin{figure*}[t!]
    \centering
    \includegraphics[width=0.95\textwidth]{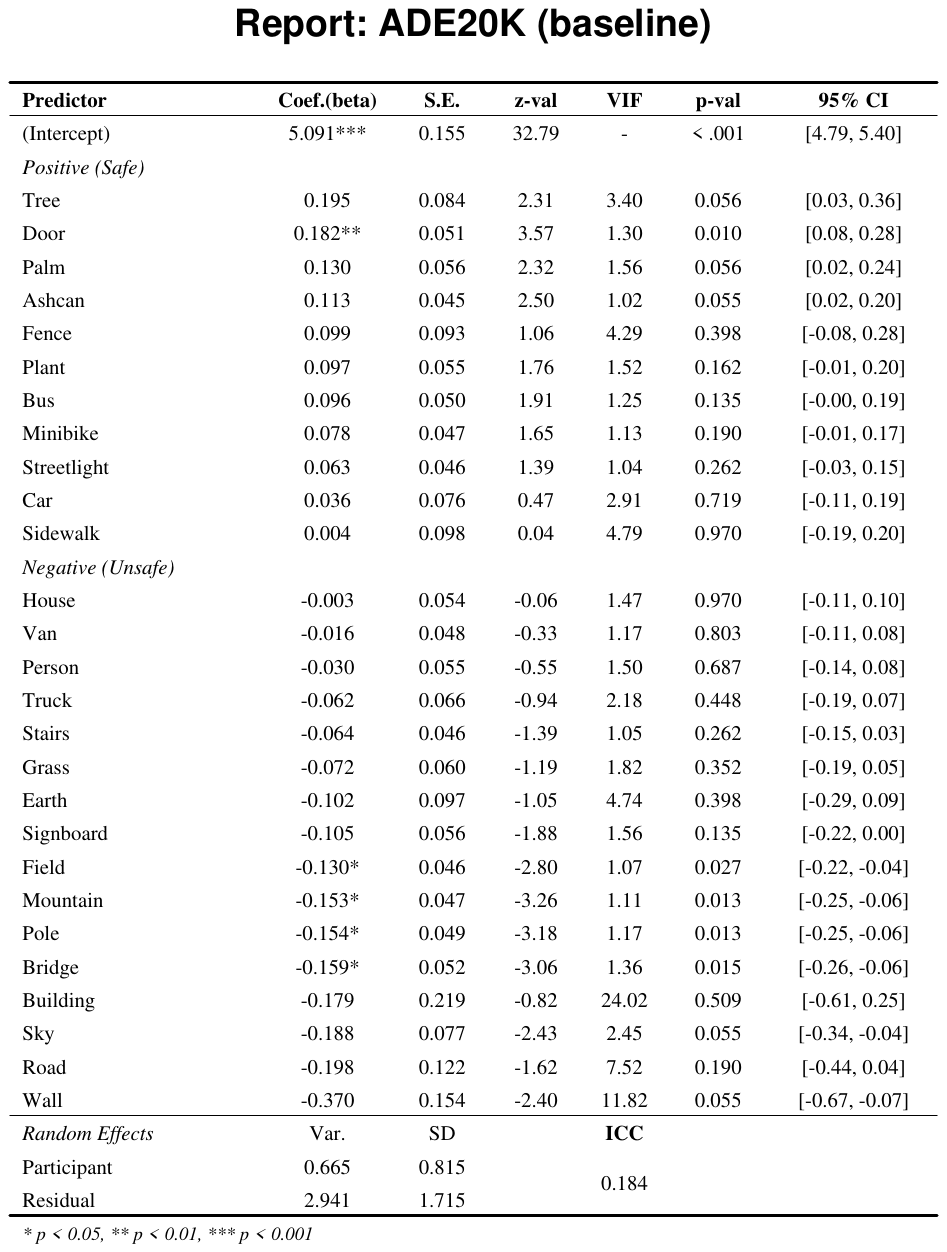} 
    \caption{Detailed statistical results for the \textit{ADE20K Base} model. The table reports the fixed effects of visual attention on perceived safety using the standard ADE20K taxonomy as a baseline.}
    \vspace{-0.4cm}
    \label{fig:lmm_base_full}
\end{figure*}
}

\newcommand{\figLmmGroupedFull}{
\begin{figure*}[t!]
    \centering
    \includegraphics[width=0.95\textwidth]{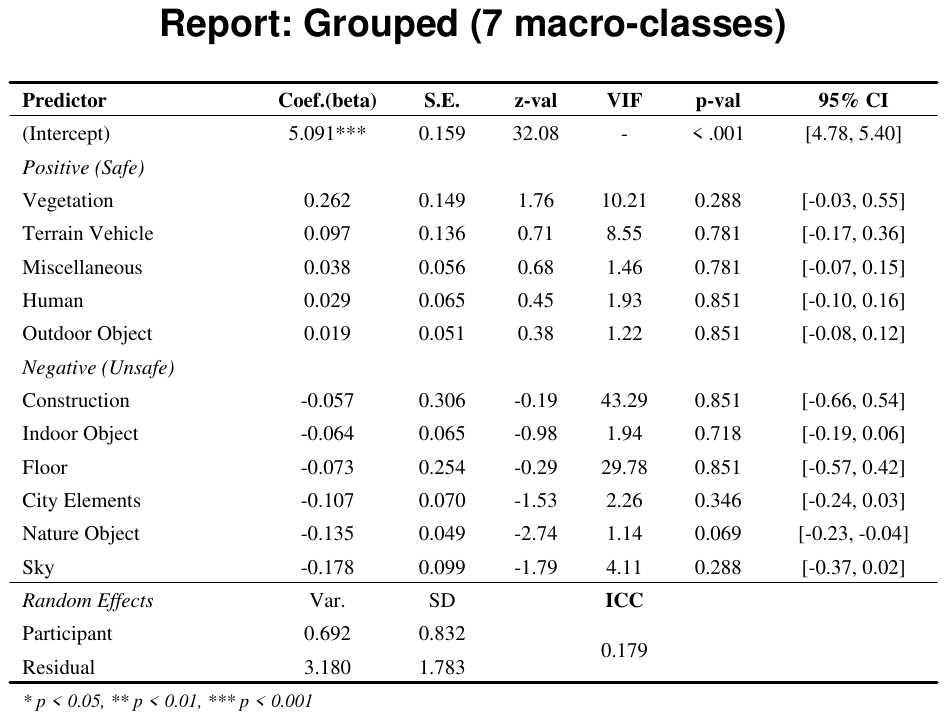} 
    \caption{Detailed statistical results for the \textit{Grouped} model. The table summarizes the effects of macro-level urban categories on perceived safety and highlights the multicollinearity introduced by broad clusters.}
    \vspace{-0.4cm}
    \label{fig:lmm_grouped_full}
\end{figure*}
}

\newcommand{\figLmmDisorderFull}{
\begin{figure*}[t!]
    \centering
    \includegraphics[width=0.95\textwidth]{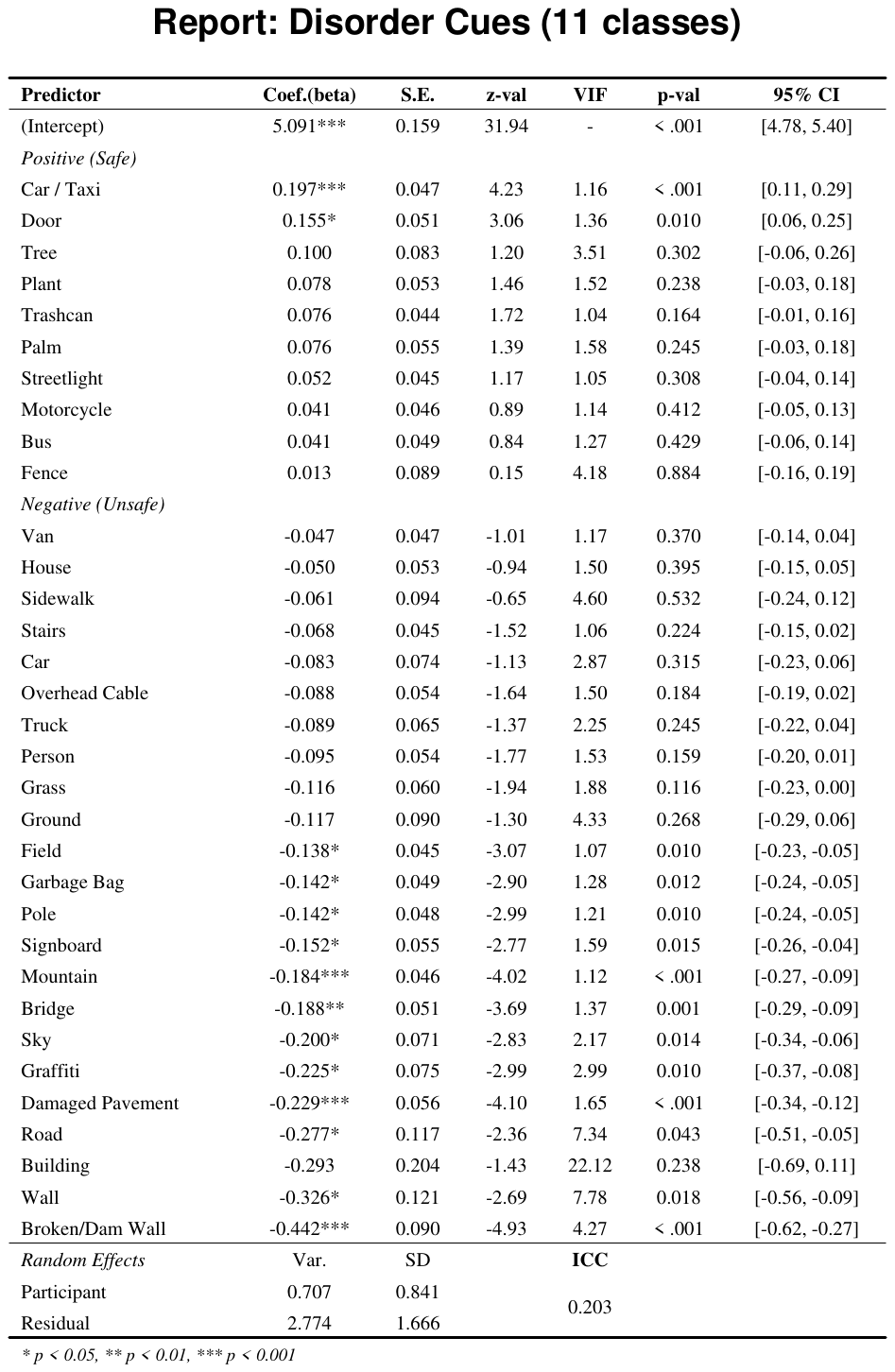} 
    \caption{Detailed statistical results for the \textit{Disorder Cues} model. The table highlights the contribution of physical disorder indicators, including graffiti and damaged structures, to perceived safety.}
    \vspace{-0.4cm}
    \label{fig:lmm_disorder_full}
\end{figure*}
}

\newcommand{\figLmmGroupedDisorderFull}{
\begin{figure*}[t!]
    \centering
    \includegraphics[width=0.95\textwidth]{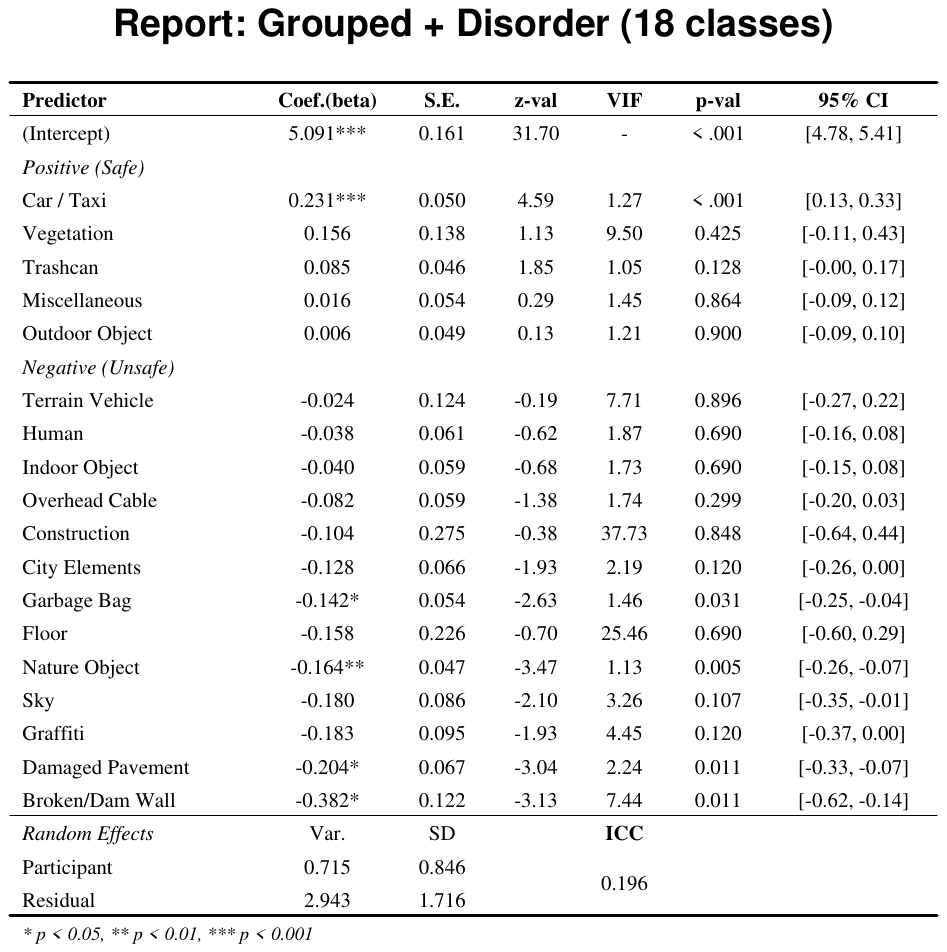} 
    \caption{Detailed statistical results for the \textit{Grouped + Disorder} model. The table evaluates the predictive contribution of disorder cues when combined with grouped urban infrastructure categories.}
    \vspace{-0.4cm}
    \label{fig:lmm_grouped_disorder_full}
\end{figure*}
}

\newcommand{\figSegmentations}{
\begin{figure}[htpb]
\centering
\includegraphics[width=\columnwidth]{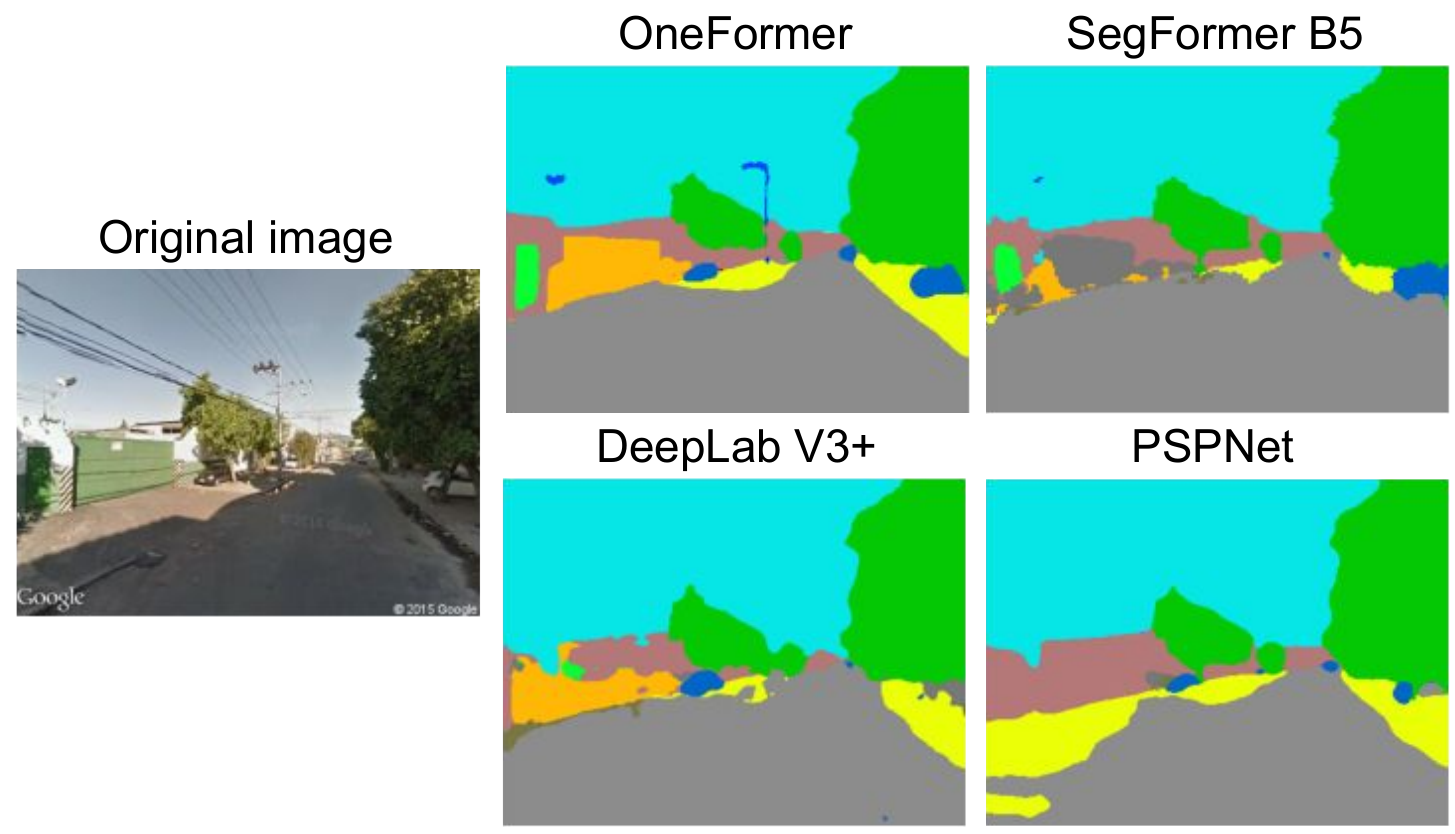}
\caption{Comparison of segmentation outputs produced by OneFormer, SegFormer B5, PSPNet, and DeepLabV3+. Some visual elements, such as light poles, cars, gates, and walls, are incorrectly classified, illustrating the differences in spatial accuracy across models.}
    \vspace{-0.4cm}
\label{fig:segmentation_model}
\end{figure}
}

\newcommand{\figTeaserDataset}{
\begin{figure}[t!]
    \centering
    \includegraphics[width=\columnwidth]{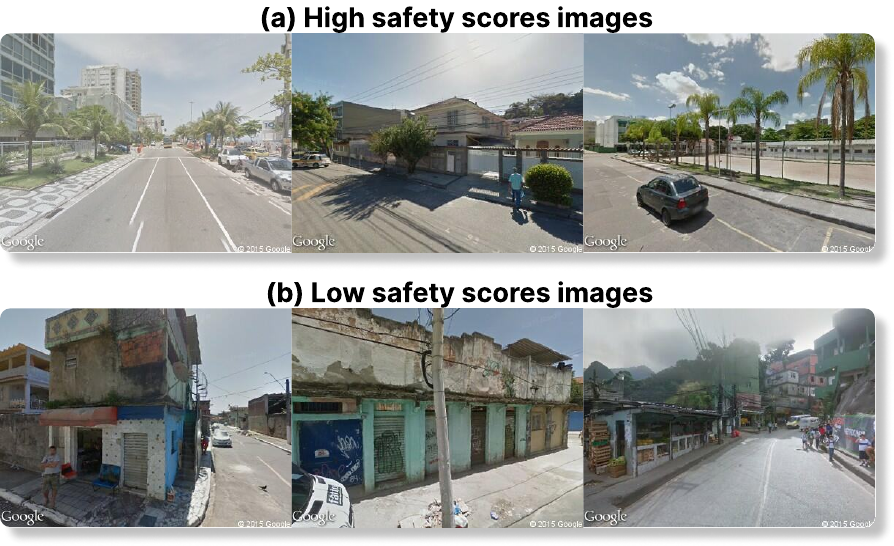}
    \caption{\textbf{Visual and structural heterogeneity within the \dataname dataset.} These representative Street View panoramas from Rio de Janeiro illustrate the contrasting urban stimuli evaluated in our study. \textbf{(a)} Well-maintained environments characterized by orderly infrastructure, clean sidewalks, and preserved vegetation, providing examples associated with higher perceived safety. \textbf{(b)} Streetscapes exhibiting pronounced physical disorder, including degraded façades, uncollected litter, and disorganized overhead cables.}
    \vspace{-0.4cm}
    \label{fig:teaser_dataset}
\end{figure}
}

\newcommand{\tabTASKS}{
\begin{table}[t!]
    \centering
    \caption{\systemname{} views and the analytical tasks they support.}
    \label{tab:urbanGazeVis_tasks}
    \begin{tabular}{llccc}
    \toprule
    Analysis & View & \textit{\textbf{T1}} & \textit{\textbf{T2}} & \textit{\textbf{T3}} \\
    \midrule
    & Image View      & \checkmark & \checkmark &            \\
    \rowcolor[HTML]{D7D7D7}
    By-Image & Attention View & \checkmark &            &  \checkmark \\
    & Temporal View            &  & \checkmark & \checkmark \\
    \midrule
    \rowcolor[HTML]{D7D7D7}
    & Attention View           & \checkmark &  & \checkmark \\
    By-Participant & Similarity View      &            &        & \checkmark \\
    \rowcolor[HTML]{D7D7D7}
    & Coverage View          &            & \checkmark  & \checkmark \\
    \bottomrule
    \end{tabular}
    \vspace{-15pt}
\end{table}
}

\newcommand{\tabICET}{
\begin{table}[t]
\centering
\caption{Average participant scores across the four ICE-T-inspired dimensions, aggregated over 22 participants using a 7-point Likert scale.}
\label{tab:icet_table}
\begin{tabular}{lcccc}
\toprule
Participant & \textit{\textbf{Insight}} & \textit{\textbf{Confidence}} & \textit{\textbf{Essence}} & \textit{\textbf{Time}} \\
\midrule
\textbf{Avg.} 
& \cellcolor{yellow!40}\textbf{5.9}
& \cellcolor{green!20}\textbf{6.3}
& \cellcolor{green!20}\textbf{6.2}
& \cellcolor{green!20}\textbf{6.1} \\
\bottomrule
\end{tabular}
\vspace{-15pt}
\end{table}
}
\section{Introduction}

Perceived safety in streetscapes is strongly shaped by visual attention, yet the relationship between gaze behavior and urban disorder remains poorly understood. Broken Windows Theory~\cite{wilson1982Broken} and subsequent empirical work~\cite{sampson2002assessing,Keizer2008Spread} suggest that visible signs of neglect, such as deteriorated façades and damaged infrastructure, reduce perceived safety and reinforce cycles of disorder, with downstream effects on behavior, crime, and mental health. Yet most approaches model safety at the image level, even though recent studies~\cite{moreno2025UrbanPD4k} suggest that individual disorder cues, such as overhead cables or structural damage, may affect perception differently.
Real-world streetscapes are also structurally and socially heterogeneous: as \autoref{fig:teaser_dataset} shows, some areas feature well-maintained, orderly infrastructure (\autoref{fig:teaser_dataset}.a), while others show pronounced disorder, such as degraded walls, litter, and irregular architecture (\autoref{fig:teaser_dataset}.b). Understanding how attention navigates these conflicting cues requires moving beyond static, image-level safety scores toward a spatiotemporal visual analytics approach.
Street View Imagery (SVI) has enabled large-scale safety studies pairing street scenes with crowdsourced judgments, as in StreetScore~\cite{naik2014streetscore}, Scenic-or-Not~\cite{Seresinhe2017UsingDL}, and Place Pulse~\cite{Dubey2016Deep}. However, these approaches mainly capture overall safety or aesthetic appeal, treating scenes holistically rather than revealing where people look or how attention unfolds over time during evaluation~\cite{Zhang2024UrbanVI,moreno2025Assessing,lavi202217k}. Recent immersive and eye-tracking studies begin to address this: virtual and mixed-reality setups allow controlled study of factors like lighting, greenery, and density in relation to safety, beauty, and liveliness~\cite{Li2022MeasuringVW}, and eye-tracking in urban contexts~\cite{Yang2024UrbanPB} identifies regions that draw attention. Still, results are typically summarized with static, aggregated fixation maps, obscuring the dynamic, temporal, and semantic dimensions of attention.

\figTeaserDataset

In this work, we adopt a visualization-first perspective. We conduct a head-mounted eye-tracking experiment using 150 street-view scenes from Rio de Janeiro, which participants rate in terms of perceived safety. We use a head-mounted optical see-through (OST) headset as a self-contained apparatus for controlled 2D stimulus presentation and gaze capture. Gaze traces are projected onto semantically segmented scenes (e.g., buildings, sidewalks, and trees) and annotated with disorder cues, including deteriorated walls, overhead cables, garbage, and graffiti. This produces a multimodal dataset integrating scene content, visual attention patterns, and perceived safety ratings.
To analyze these data, we introduce \systemname, a visualization system for exploring eye-tracking data in urban safety studies. \systemname supports post hoc inspection of spatiotemporal gaze behavior in relation to semantic regions and safety judgments, linking gaze trajectories, fixation patterns, and scene elements to support both hypothesis generation and evidence assessment for researchers and practitioners. Our contributions are threefold:

\myparagraph{Head-mounted eye-gaze collection methodology.} A head-mounted eye-tracking protocol for selecting, filtering, and assigning SVI, including calibration, trial design, and quality control, to collect curated safety judgments.

\myparagraph{\dataname dataset.} A multimodal dataset of Rio de Janeiro street-view images with safety ratings, eye-tracking data, semantic segmentations, and participant metadata for urban perception research.

\myparagraph{\systemname visualization system.} An interactive system that overlays dynamic gaze information on semantically segmented scenes, enabling exploration of gaze paths and fixation patterns in relation to safety scores through domain-inspired case studies.

All datasets, supplemental materials, and source code will be made available upon acceptance, in accordance with the venue's anonymization guidelines.
\section{Related Work}
This research lies at the intersection of five pillars: (i) Urban Perception and Computer Vision, (ii) Environmental Criminology,  (iii) Eye-Tracking for Urban Perception, (iv) Immersive and Mixed Reality in Urban Visualization, and (v) Visual Analytics for Eye-Tracking Data.

\myparagraph{Urban Perception and Computer Vision.}
Urban perception research examines how the built environment is visually characterized and how people judge properties such as safety from street-level imagery~\cite{Santani2018LookingS,moreno2021QuantifyUrbanPerception,moreno2021UrbanPercetion}. Large-scale image collections paired with crowdsourced judgments, such as Place Pulse and StreetScore~\cite{Salesses2013TheCI,naik2014streetscore}, support questions such as ``\textit{What makes Paris look like Paris?}''~\cite{Doersch2012WhatMP} and ``\textit{What makes a place feel safe?}''~\cite{moreno2024WhatMakes}. Early work relied on hand-crafted features; later approaches adopted CNNs to predict perceived attributes~\cite{Ordonez2014LearningHJ,Porzi2015PredictingAU,Dubey2016Deep}, and more recent methods relate semantic scene content to safety and other judgments~\cite{Zhang2018MeasuringHP,Min2020MultiTaskDR}, bridging perception studies and visual analysis. Most of these approaches treat perception as an \emph{image-level} label, learning from global ratings without observing how viewers explore a scene. Our work extends this line by linking safety judgments to \emph{where} observers look, using semantic classes and disorder cues in an interactive visual analytics workflow.

\myparagraph {Environmental Criminology.  }
While computer vision frameworks traditionally model urban perception through passive image-level feature extraction, environmental criminology establishes safety perception as a dynamic cognitive tension between physical infrastructure and social activity. 
This perspective draws on foundational frameworks in environmental criminology---
routine activity theory~\cite{cohen1979social}, crime pattern
theory~\cite{brantingham1993environment}, and situational crime
prevention~\cite{clarke1983situational}---which we invoke not as models of crime occurrence
but as an interpretive lens for how everyday spatial routines and the configuration of the
built environment shape where people perceive safety or threat. Jane Jacobs's ``eyes on the street''~\cite{jacobs1961death} and Oscar Newman's ``defensible space''~\cite{newman1972defensible} further assert that human presence and street vitality fundamentally alter how an environment is monitored and interpreted, principles underlying Crime Prevention Through Environmental Design (CPTED), which identifies ``natural surveillance'' as a primary psychological mitigator of perceived threat. Rather than treating pedestrians, crowds, and street activity as mere geometric or baseline semantic classes, our approach conceptualizes human presence as an active social cue that can dynamically modulate how observers interpret co-occurring signals of physical disorder.

\myparagraph{Eye-Tracking for Urban Perception.}
Eye-tracking methodologies offer a window into spatial cognition, decomposing visual attention into fixations, saccades, and scanpaths~\cite{Yang2024UrbanPB,Yang2024UrbanPA}. 


Desktop and head-mounted eye-tracking systems involve different trade-offs in
sampling rate, head-movement tolerance, calibration, and stimulus presentation, and the
appropriate choice depends on the study design ~\cite{Li2022MeasuringVW,Yang2024UrbanPB} .
As synthesized by Moreno-Arjonilla et al.~\cite{moreno2024survey}, integrating head-mounted eye trackers within spatial computing frameworks demands rigorous calibration and geometric uncertainty modeling to yield reproducible attentional datasets. Contemporary work increasingly intersects spatiotemporal gaze with deep feature spaces and explainable AI to isolate visual correlates of safety and aesthetic judgment~\cite{kang2026decoding,Oki2021EvaluatingVI}. Although these studies show gaze provides rich information about safety judgments, results are usually presented as static heatmaps, aggregate statistics, or a few example scanpaths—summaries that flatten a dynamic process and obscure how attention shifts over time and across semantic elements. Our work treats the HoloLens and calibration as infrastructure, focusing on structuring and visualizing spatiotemporal gaze data through a safety-judgment protocol and \systemname for multimodal exploration.


\myparagraph{Immersive and Mixed Reality in Spatial Visualization.}
The convergence of spatial computing and visual analytics has shifted data exploration from desktop interfaces toward immersive environments. Recent advances emphasize situated visualizations, where geometric data and semantic layouts are embedded within the user's egocentric frame to enhance cognitive processing and environmental presence~\cite{boorboor2023submerse, perez2024using}. Immersive analytics has evolved to leverage advanced view computation and cross-scale interaction, letting analysts explore large-scale urban representations and complex environmental phenomena fluidly~\cite{cobeli2025neural, ouyang2025oceanvive}. Furthermore, contemporary frameworks integrate realistic 3D modeling with interactive simulations to evaluate spatial contexts and urban risks in real time~\cite{banno2025360citygml, gonzalez2025floodsim}. 

 \systemname{} sits within this frontier but with a deliberately narrow scope. We do not use Mixed Reality (MR) for situated 3D visualization; the headset serves only as a self-contained eye tracker that presents 2D SVI and records gaze. Our contribution is not the display medium but the protocol and visual analytics that relate the resulting spatiotemporal gaze data to semantic scene content and fine-grained urban decay.

\myparagraph{Visual Analytics for Eye-Tracking Data.}
Eye-tracking visualization is a mature research vector within visual analytics. Established studies~\cite{blascheck2017visualization,Andrienko2012VAMethodology} categorize techniques across spatial, temporal, and hybrid spatiotemporal dimensions, using in-context gaze overlays, timelines, and coordinated views to reduce cognitive overhead. Early paradigms such as space--time cubes~\cite{Kurzhals2013SpacetimeVA,Heimerl2014ISeeCube} and AOI-centric encodings like gaze stripes and rivers~\cite{Burch2013AOIRivers,Kurzhals2016GazeStripes,Morimoto2015HeatmapExplorer} optimized localized sequence extraction. More recently, the VETA system~\cite{goodwin2022veta} provides diagnostic interfaces for spatial gaze distributions, while SVATA~\cite{ren2026svata} introduces infrastructure for multi-user fixation analysis, implementing an Area-Fixation Weight (AFW) metric that balances spatial distribution bias—conceptually mirroring the area-normalized attention intensity metric (AttIn) we propose in \autoref{sec:attention_intensity_metric}. Both frameworks confront spatial dominance artifacts by discount-weighting raw fixation durations relative to bounding visual area, emphasizing genuinely salient features over structurally dominant background regions. 

Building on these ideas, \systemname targets a different setting: rather than abstract AOIs or screen regions, we project gaze onto semantic segments (e.g., buildings, greenery, sky) and urban disorder cues (e.g., damaged façades, exposed cables) in SVI, linking these encodings to time-resolved views and safety scores. This lets analysts explore \emph{which} urban elements attract attention, \emph{when} they are inspected, and \emph{how} patterns differ between scenes perceived as safe or unsafe—adapting general eye-tracking visualization concepts to a domain-specific, semantics-aware analysis of urban streetscapes.
\section{System Overview} 
\label{section:system_overview}

\figWorkflow

Eye-tracking research shows that gaze data are high-volume and spatiotemporal, yet perceptual judgments are often reduced to image-level scores, leaving interactive exploration of gaze--semantics--rating relationships limited~\cite{blascheck2017visualization,Andrienko2012VAMethodology}. Drawing on ``overview first, zoom and filter, details-on-demand''~\cite{shneiderman2003eyes} and established gaze-analysis task abstractions~\cite{blascheck2017visualization}, we formulated \systemname's design goals and analytical tasks.

\subsection{Design Goals}

Informed by eye-tracking taxonomies~\cite{blascheck2017visualization}, spatiotemporal visual analytics~\cite{Andrienko2012VAMethodology}, and urban perception studies interpretability~\cite{Zhang2018MeasuringHP,Min2020MultiTaskDR}, our design goals are: \textbf{DG1: Support image- and participant-centric analysis} of how multiple individuals explore one streetscape and how one individual behaves across scenes, using consistent encodings~\cite{blascheck2017visualization}; \textbf{DG2: Characterize spatiotemporal gaze behavior}, revealing temporal dynamics and scanpath differences hidden by static heatmaps and aggregates~\cite{Andrienko2012VAMethodology}; and \textbf{DG3: Relate gaze, semantics, and safety judgments} by linking gaze traces to semantic segments, disorder cues, and ratings~\cite{Zhang2018MeasuringHP,Min2020MultiTaskDR,moreno2024WhatMakes}. These goals motivate dual views (\emph{By-Image} and \emph{By-Participant}) with coordinated summaries of spatial, temporal, and semantic gaze patterns.

\subsection{Analytical Tasks} \label{sec:analytical_tasks}

Grounded in these goals and eye-tracking task taxonomies~\cite{blascheck2017visualization}, we define analytical tasks following an overview-to-detail pattern~\cite{shneiderman2003eyes}: from a global summary of gaze and ratings, to inspecting images or participants, to detailed traces:

\myparagraph{T1: Spatial attention analysis (DG1, DG2)} summarizes which regions and semantic elements attract attention across participants for a selected image (\textbf{T1.1}), and where a selected participant tends to look across all assessed images (\textbf{T1.2}). 

\myparagraph{T2: Temporal gaze pattern exploration (DG1, DG2)} visualizes how attention moves across an image over time to compare scanpaths between participants (\textbf{T2.1}), and how a participant's gaze evolves across a session to detect learning or fatigue effects (\textbf{T2.2}). 

\myparagraph{T3: Linking gaze, semantics, and safety assessments (DG3)} relates fixations on semantic categories and disorder cues to the safety rating for a specific image (\textbf{T3.1}), and compares a participant's semantic focus and gaze coverage with their score distribution to inspect rating consistency across visually similar images (\textbf{T3.2}). 

These tasks inform \systemname's two main components: a \emph{By-Image} view for image-centric tasks (\textbf{T1.1}, \textbf{T2.1}, \textbf{T3.1}) and a \emph{By-Participant} view for participant-centric tasks (\textbf{T1.2}, \textbf{T2.2}, \textbf{T3.2}), detailed in \autoref{section:urbangazevis}.

\section{\dataname Dataset}
\label{section:perception_assessments}

\autoref{fig:workflow} summarizes our end-to-end workflow, from multimodal dataset curation using SVI and head-mounted eye-tracking to the interactive visual analytics environment detailed in \autoref{section:urbangazevis}. This section covers the first phase: constructing \dataname.

\subsection{Image Curation and Semantic Extraction}
\label{section:image_curation}

We curated a compact but diverse set of streetscapes, distributed evenly across participants to support the analyses enabled by \systemname.
We used the Urban Physical Disorder-4k (UrbanPD4k) dataset~\cite{moreno2025UrbanPD4k}, which contains Rio de Janeiro SVIs with pixel-level annotations of physical disorder cues. We applied the OneFormer~\cite{jain2023oneformer} model pre-trained on ADE20K dataset to obtain urban elements. To support fine-grained analysis in \systemname, we define four cases based on data:

\myparagraph{\textbf{ADE20K (baseline):}} 150 segmentation classes (e.g., car, building, tree, road).

\myparagraph{\textbf{Grouped-ADE (7 macro-classes):}} 7 segmentation group classes: \textit{City Elements, Construction, Floor, Human, Sky, Terrain Vehicle,} and \textit{Vegetation}.

\myparagraph{\textbf{ADE20K+UrbanPD4k:}} Adds physical disorder classes: \textit{Overhead cables, Graffiti, Broken/Damaged Bricks Wall, Damaged Pavement/Road, Car Taxi, Damaged Traffic Sign, Garbage Bag, Garbage Box, Kiosk, Exposed Overhead Cables,} and \textit{Trashcan}.

\myparagraph{\textbf{Grouped-ADE + UrbanPD4k:}} Groups and disorder classes.

\subsection{Multivariate Stratified Sampling}

\label{sec:sampling}
We used \textit{multivariate stratified sampling} to select a subset suitable for a 
controlled experiment while preserving the visual diversity of the full dataset. Rather than manually normalizing for varying traffic or pedestrian counts, we formulated the selection as an Integer Linear Programming (ILP) problem over the baseline ADE20K segmentation layouts. For each structural and social class (e.g., buildings, roads, vegetation, cars, pedestrians), pixel-ratio distributions were partitioned into quartiles, and the ILP solved for a balanced sample of 150 images with equal representation across all baseline strata (e.g., vehicles, pedestrians, vegetation), though it does not model all acquisition conditions such as time of day. This balancing mitigates the risk of any single layout category acting as an uncontrolled confound, giving our post-hoc regression models the statistical stability to isolate the perceptual weights of general infrastructure and co-occurring disorder features. Mathematical details of the ILP objective and constraints are in~\ref{appendix:math_form}.

\subsection{Image Assignment Protocol}

\label{sec:assignment_protocol}
To balance the experimental design without participant fatigue, the 150 images were distributed via randomized block assignment. The 30 participants were split into three blocks of 10 raters each, with each block evaluating a unique subset of 50 streetscapes drawn from the ILP strata. Within each evaluation session, these 50 images were presented to the participant in a completely randomized sequence. This guarantees every image was evaluated by exactly 10 distinct participants, mitigating systematic presentation-order bias while maintaining structural symmetry across the sparse measurement matrix (full matrix in~\ref{appendix:random_matrix}).

\figImageAnalyzer

\subsection{Participant Study and Data Collection}
To build \dataname, we conducted a 
controlled experiment in which participants rated the safety of SVI while their eye movements were recorded.

\myparagraph{Participant demographics and inclusion criteria.}
We recruited 30 volunteers via institutional mailing lists, with informed consent obtained prior to participation. The cohort included 16 undergraduate, 9 master, and 5 doctoral students in fields such as urban planning, computer science, and social sciences. Twenty participants were Brazilian (from seven states), seven Peruvian, and three Paraguayan. Ages ranged from 19 to 43 (median $= 23.8$, SD $= 5.2$), with 20 men and 10 women. Participants with conditions incompatible with head-mounted eye tracking were excluded. In total, the study yielded 1,500 trials.
We used a Microsoft HoloLens~2 solely as a head-mounted eye tracker, chosen to integrate calibration, stimulus presentation, and gaze capture in one setup. Each SVI was shown as a flat 2D stimulus on a frontoparallel virtual plane at 2.0~m (800$\times$600~px); using the standard Eye Tracking API~\cite{microsoft_eyetracking}, we recorded the single eye-gaze ray (origin and direction) at approximately 30~Hz and intersected it with this plane to obtain 2D coordinates. Dedicated desktop trackers offer higher sampling rates and remain viable alternatives; we account for the resulting spatial uncertainty through the smoothing model in \autoref{section:dataset_validation}. Each session began with a nine-point gaze calibration and a brief familiarization phase with reference images. During main trials, participants viewed each image for
15 seconds while gaze data were recorded continuously, then rated perceived safety on a 1--10 Likert scale via a handheld controller.
Sessions lasted about 25 minutes; additional protocol details are
in~\ref{appx:experimental_protocol}.

\myparagraph{Experimental Controls and Noise Mitigation.}
All sessions were conducted in a dedicated, climate-controlled laboratory. Participants faced a blank, matte-white wall, and stimuli were rendered at maximum opacity so that the physical background could not show through or trigger exogenous saccades and peripheral distractions.
Each entry in the resulting \dataname dataset includes:
\begin{itemize}[noitemsep]
\item \textbf{A street-view image} and its geographic coordinates (latitude and longitude);
\item \textbf{A temporal sequence of gaze points} $(x, y, t)$ recorded during the 15-seconds observation interval;
\item \textbf{A Likert-scale safety rating (1--10)}, where 1 denotes ``very unsafe'' and 10 denotes ``very safe''.
\end{itemize}
\section{Dataset Validation and Robustness}
\label{section:dataset_validation}

\figParticipantAnalyzer

Before integrating the collected multimodal data into \systemname, we conducted a quantitative validation of the dataset, assessing signal stability, inter-rater reliability, and structural validity.

\subsection{Hardware Robustness and Spatial Uncertainty}
Robustness was first addressed at the hardware level to minimize coordinate mapping uncertainty and segmentation-boundary noise. Due to the inherent physiological and hardware jitter of head-mounted eye tracking, raw fixation points were modeled as attentional catchment areas rather than absolute pixel coordinates. We applied a Gaussian smoothing kernel with a standard deviation of $\sigma = 10.68$ pixels. This smoothing parameter was not empirically tuned; it was analytically derived from the Microsoft HoloLens 2 empirical angular precision ($0.24^\circ$) and the effective pixel-per-degree resolution of our display setup to map hardware uncertainty into the stimulus space (see \ref{appx:Kernel_Calculation} for the geometric derivation). This provides a reliable foundation for evaluating attention on small, complex urban cues.

\subsection{Inter-Rater Reliability Analysis}
To evaluate the consistency of urban safety perceptions across participants, we performed a rater calibration analysis tailored for sparse rating matrices. Given our Ill-Structured Measurement Design (ISMD)—where 30 participants evaluated overlapping but incomplete subsets of the 150 images—traditional Intraclass Correlation Coefficients (ICC) based on balanced ANOVA are mathematically biased.

Following Putka et al.~\cite{putka2008ill}, we used a Linear Mixed-Effects Model (LMM) with crossed random effects to isolate the variance components of the $1,500$ observations, partitioning total variance into the urban scene (signal), the participant (subjective bias), and residual noise. The model-based reliability coefficient $G(q,k)$ was computed as follows:
\begin{equation}
G(q,k) = \frac{\sigma_{\text{img}}^2}{\sigma_{\text{img}}^2 + \frac{q \, \sigma_{\text{rat}}^2 + \sigma_{\text{res}}^2}{k}}
\label{eq:putka_reliability}
\end{equation}

Where $k = 10$ is the number of raters per image, and $q \approx 0.67$ is the coefficient of non-overlap derived from our sparse matrix layout ($q = 1 - k/N_{\text{total}}$), reflecting that each image was evaluated by one-third of the pool. The analysis yielded a high inter-rater reliability of $G(q,k) = 0.812$. In behavioral research, values above $0.80$ indicate robust consensus, indicating that 10 raters per image provide sufficient reliability to capture collective perception in this study.

\subsection{Baseline Variance Decomposition}
To validate the dataset structure, a baseline LMM was fitted with participants as random effects to account for variance in baseline severity or leniency. This model achieved an ICC of $0.203$, indicating that approximately one-fifth of the variance in safety scores is attributable to participant-specific baselines, with the remainder attributable to image-level differences and residual factors. These findings confirm that while safety perception has a subjective component, a substantial portion of the variation is linked to image-level differences, justifying the development of an interactive visual analytics platform to isolate these signals.
\section{UrbanGazeVis} 
\label{section:urbangazevis}

\systemname is an interactive visual analytics system addressing the tasks outlined in \autoref{section:system_overview}, offering two coordinated analysis modes: \emph{By-Image} and \emph{By-Participant}. These modes let analysts alternate between image- and participant-centric perspectives (\textbf{DG1}), characterize spatiotemporal gaze behavior (\textbf{DG2}), and relate gaze patterns to semantic content and safety ratings (\textbf{DG3}); \autoref{tab:urbanGazeVis_tasks} summarizes how each view supports tasks \textbf{T1--T3}.
\subsection{Data Abstraction and Preprocessing}
\label{sec:data_abstraction}
Raw eye-tracking data form a continuous stream of gaze coordinates comprising fixations and saccades. Since visual information is primarily acquired during fixations, saccadic samples are excluded prior to analysis using the Identification by Velocity Threshold (I-VT) algorithm~\cite{SalvucciGoldberg2000}: velocity is computed between consecutive samples, and sequences below $1.15 \text{ px/ms}$ are classified as fixations, consistent with established protocols for head-mounted displays~\cite{karthik2019custom}, while samples exceeding this threshold are labeled saccades and excluded.
To establish a baseline for \textit{Saliency Coverage}, we construct Empirical Saliency Maps (ESMs) from aggregated human gaze rather than computational saliency models. Gaze points are smoothed with a Gaussian kernel ($\sigma=10.68$ pixels), analytically derived from the HoloLens 2 empirical angular precision ($0.24^\circ$) and screen resolution specifications to map hardware uncertainty into the stimulus pixel space (see \ref{appx:Kernel_Calculation} for the geometric derivation).

Spatial dispersion of attention is quantified using \textit{spatial entropy}, defined as Shannon entropy over normalized fixation density~\cite{shannon1948mathematical}, a standard measure of gaze dispersion~\cite{le2013methods,borji2012state}. Fixations are then mapped onto semantic segmentations to estimate \textit{Attention Intensity} for each urban class.

\tabTASKS

\figScenarioA

\subsection{Attention Intensity Metric}
\label{sec:attin_metric} \label{sec:attention_intensity_metric}
To examine the relationship between the built environment and safety perception, we quantify the attention participants allocate to each visual element~\cite{moreno2025UrbanPD4k}. Total attention is computed as accumulated fixation time; however, large objects (\eg, façades or sky) naturally receive more fixations due to their greater pixel area.

To reduce this bias, we introduce an area-normalized attention intensity ($AttIn$) that discounts each class by its image area, where $t_{p,i,c}$ denotes the total time (ms) participant $p$ fixated on class $c$ in image $i$, and $r_{i,c} \in (0,1]$ is the fraction of pixels belonging to class $c$.

To avoid instability when normalizing by very small regions ($r_{i,c} \rightarrow 0$), a lower-bound threshold $\tau$ is applied; given the eye tracker's spatial uncertainty, $\tau$ is set to $0.005$ ($0.5\%$), and regions below this threshold are excluded as indistinguishable from hardware noise (detailed formulation in \ref{appx:Kernel_Calculation}). The resulting metric is defined as follows:

\begin{equation}
\label{eq:attin_formula}
AttIn_{i,c} = \frac{1}{P_{i}} \sum_{p=1}^{P_{i}} \frac{t_{p,i,c}}{\max(r_{i,c}, \tau)}
\end{equation}

This formulation highlights genuinely salient small objects, such as graffiti or cables, while maintaining robustness to segmentation noise.

\figScenarioB

\subsection{By-Image Component}
The By-Image component supports image-centric analysis, visualizing how participants explore a selected streetscape via a coordinated control panel and four views (\autoref{fig:image_analyzer_component}). The \textbf{Control Panel} (\autoref{fig:image_analyzer_component}a) enables image selection, sorting by safety scores, participant filtering, and data toggling (e.g., ADE20K or disorder cues) to coordinate linked views (\textbf{T1--T3}). The central \textbf{Image View} (\autoref{fig:image_analyzer_component}b) shows the selected streetscape with optional overlays for raw gaze, density contours, or heatmaps, and toggles between original and semantic representations (\textbf{T1}). For detailed ROI-centered analysis, a circular Region of Interest (ROI) is instantiated via a single click; analysts can drag and scale this selector to trigger the \textbf{Glyph View} (\autoref{fig:image_analyzer_component}c), summarizing visual exploration within the designated bounds over time (\textbf{T2}, \textbf{T3}).
The Glyph View's ring functions as a 15-second clock: stacked radial bars represent participant counts per 1-second segment, revealing early versus sustained attention. Hovering over a segment reports the participant IDs contributing to that interval; filtering the Control Panel to a single participant turns the glyph into an individual 15-second timeline, making sustained attention and revisits explicit. The \textbf{Attention View} (Participant Attention Summary, \autoref{fig:image_analyzer_component}d) relates participants (columns) to semantic classes (rows), encoding total time or our AttIn metric (\autoref{eq:attin_formula}) via heatmap intensity. Finally, the bottom \textbf{Temporal View} (Temporal Participant Attention, \autoref{fig:image_analyzer_component}e) uses a scarf-timeline chart, where rows represent participants and colored segments encode semantic class fixations (\textbf{T2}, \textbf{T3}), revealing scanpath diversity and transitions between urban elements.

\subsection{By-Participant Component}
This component offers a complementary perspective, aggregating behavior across images viewed by a selected participant (\autoref{fig:participant_analyzer_component}). Its \textbf{Control Panel} (\autoref{fig:participant_analyzer_component}a) specifies attention types and data, including normalization toggles for the attention heatmap. The participant-level \textbf{Attention View} (\autoref{fig:participant_analyzer_component}b) transposes the previous heatmap: columns represent images (sorted by safety score) and rows represent semantic classes (\textbf{T1}, \textbf{T3}). The \textbf{Similarity View} (Image Projection, \autoref{fig:participant_analyzer_component}c) shows a 2D t-SNE projection of the 50 images viewed by the participant, using 4096-dimensional Places365 embeddings~\cite{zhou2017places}, colored by safety score to facilitate comparisons across visually similar scenes (\textbf{T3}); the \textbf{Image View} (\autoref{fig:participant_analyzer_component}d) displays the streetscapes selected from that projection alongside their safety scores. Finally, the \textbf{Coverage View} (Saliency coverage comparison, \autoref{fig:participant_analyzer_component}e) relates \textit{Saliency Coverage} or \textit{Stationary Entropy} to safety scores or image order (\textbf{T2}, \textbf{T3}), letting analysts assess whether perceived safety correlates with broader versus focused exploratory visual strategies.

\subsection{Implementation Details}

\systemname{} employs a decoupled client--server architecture: client-side coordinated views are implemented in JavaScript with D3.js~\cite{bostock2011d3}, while the server uses Python and Flask~\cite{grinberg2018flask} for data indexing. To ensure responsive interactions, computationally expensive feature extraction, dimensionality reduction, and segmentation are performed offline.

Our preprocessing pipeline uses OneFormer~\cite{jain2023oneformer} (pretrained on ADE20K via HuggingFace~\cite{wolf2019huggingface}), selected for its superior label correctness and boundary alignment over alternatives (\eg, DeepLabv3+, SegFormer; see \ref{appendix:segmentation_model}). For similarity-driven exploration in the By-Participant component, we extract 4,096-dimensional scene embeddings with a pretrained Places365Net~\cite{zhou2017places}. Unlike raw semantic maps or general-purpose models such as CLIP, Places365Net better captures urban layouts and structural decay, preventing visually distinct but semantically similar scenes from appearing artificially identical. The embeddings are projected into 2D using \emph{t}-SNE~\cite{pedregosa2011scikit} with PCA initialization, preserving local neighborhoods and enabling stable comparisons within each participant's image subset (\(N = 50\)).

\section{Usage Scenarios}

Three usage scenarios illustrate how \systemname supports the analysis of eye-gaze data in urban safety studies.

\figScenarioC

\subsection{Associations between Safety Perceptions and the Built Environment} \label{sec:scenario_1}

\autoref{fig:scenario_1} presents the analytical workflow within the \emph{By-Participant} component for Participant~29. Three streetscapes are selected for detailed inspection, denoted by their unique dataset identifiers: Image ID 60 (\textcolor[HTML]{66C2A5}{mint green}), Image ID 38 (\textcolor[HTML]{FC8D62}{coral}), and Image ID 109 (\textcolor[HTML]{8DA0CB}{soft blue}), sorted left to right by mean perceived safety score, from unsafe to safe.
The default view displays \textit{total attention} (top heatmap), which emphasizes large background elements such as construction and floor. Switching to \textit{attention intensity} (\(\mathrm{AttIn}\), defined in \autoref{sec:attin_metric}) instead highlights elements with low spatial coverage but strong attentional focus, particularly disorder cues such as graffiti and damaged walls. In Image~60 (top row), total attention emphasizes construction, whereas \(\mathrm{AttIn}\) reveals stronger emphasis on smaller elements such as vehicles and city features like street lamps. In Image~38 (middle row), total attention highlights construction, floor, and graffiti, while \(\mathrm{AttIn}\) concentrates on graffiti, linking infrastructural decay to reduced perceived safety. In Image~109 (bottom row), \(\mathrm{AttIn}\) again emphasizes graffiti and damaged walls over the broader streetscape.
By isolating these subtle yet influential elements, \systemname demonstrates that combining \(\mathrm{AttIn}\) with disorder cues helps identify key correlates of safety perception.

\subsection{Analyzing Divergent Scoring Patterns} \label{sec:scenario_2}

Assessing how consistently participants score visually similar images reveals the stability of their judgments; large rating differences for similar scenes may indicate sensitivity to specific visual cues. Using \(\mathrm{AttIn}\), an analyst selects a participant in the \emph{By-Participant} component and uses the \textbf{Similarity View} to find nearby images in the projection space that received different scores. The \textbf{Attention View} and \textbf{Coverage View} summarize visual focus, while gaze points are overlaid in the \textbf{Image View}.

\autoref{fig:usage_case_2} presents three examples for Participants~10, 22, and 2. In \autoref{fig:usage_case_2}.a, Participant~10 rates Image~145 highly (8) but Image~100 low (2): attention in Image~145 is balanced between \textit{Construction} and \textit{Vegetation}, whereas in Image~100 dense fixations concentrate on \textit{Graffiti}, contributing to the lower score.
In \autoref{fig:usage_case_2}.b, Participant~22 rates Image~72 as 6 and Image~38 as 3. Although both scenes contain graffiti, gaze patterns differ: Image~72 attracts attention to \textit{Construction}, \textit{Graffiti}, and \textit{Human/pedestrians}, yielding a moderate score, while gaze in Image~38 stays largely confined to \textit{Construction} and \textit{Graffiti}. This suggests that while physical disorder is associated with lower perceived safety, the presence of people can partially mitigate this negative association.
Finally, \autoref{fig:usage_case_2}.c shows Participant~2's evaluations of Images~72 and~10. Despite prominent graffiti, Image~72 (score 7) draws substantial attention to pedestrians, while Image~10 (score 3), lacking human presence, directs attention toward vehicles and \textit{Broken Road} cues. Across these cases, \systemname reveals a consistent pattern: the visual presence of people often shows a stronger positive association with perceived safety than graffiti alone.

\subsection{Spatiotemporal Dynamics and Participant Divergence}
\label{sec:scenario_3}

In this scenario, we investigate the spatiotemporal dynamics of attention to understand how different observers process the same streetscape during the 15-second trial (\autoref{fig:scenario_3}). Aggregate heatmaps reveal where attention is concentrated but not whether a region was fixated continuously, intermittently, early, or late. We analyze Image ID~20 with a safety score of 4.3 (\autoref{fig:scenario_3}.a), whose segmentation identifies all semantic objects. Although the \textit{person} class (pedestrian area) occupies only a small portion of the image, the \textbf{Image View} heatmap (\autoref{fig:scenario_3}.b) highlights it as a visual hotspot.

To investigate this hotspot, the ROI Glyph links the selected region to the coordinated \textbf{Temporal View} (\autoref{fig:scenario_3}.c), which decomposes the 15-second scanpaths of individual participants. This view reveals distinct exploration strategies. Participant~27 (P-27) fixates the \textit{person} class only during the initial seconds (early single visit), whereas Participant~30 (P-30) attends to it only near the end of the trial (late single visit). In contrast, Participant~16 (P-16) repeatedly returns to the \textit{person} class after exploring other semantic elements (revisit).

Together, the ROI Glyph and \textbf{Temporal View} expose temporal behaviors—early visits, late visits, and revisits—that are not visible in conventional spatial heatmaps. These differences suggest that human presence is attended not only as a semantic category but also at different stages of the visual assessment process, providing insights into how social cues may influence perceived safety.

\section{System Evaluation and Value-Driven Assessment}
\label{section:evaluation}

A user study with 22 participants from diverse academic and professional backgrounds evaluated the analytical effectiveness and value-driven utility of \systemname. The study intentionally focused on individuals without formal expertise in eye-tracking analysis, to test whether the system makes complex visual behavior data accessible to a broader audience.

\subsection{Study Setup and Participants} 
We recruited 22 participants with experience in general data analysis but no prior exposure to eye-tracking data. The evaluation design and questionnaire were explicitly inspired by the ICE-T framework~\cite{wall2018heuristic}, which assesses visualization systems along four dimensions: Insight, Confidence, Essence, and Time. Prior research indicates that relatively small samples suffice to capture exploratory feedback and assess visualization effectiveness~\cite{wall2018heuristic}, supporting this sample size. Although participants were familiar with system evaluations, none had previously used \systemname.

\subsection{Tasks and Procedure}
\label{sec:tasks_and_procedure}

The study was conducted in person. Participants received a 10-minute introduction to the \systemname interface, covering the By-Image and By-Participant views and their primary interactions, then completed four visual analytics tasks (\textbf{U1--U4}) engineered to operationalize and cover the high-level analytical design tasks (\textbf{T1--T3}) defined in \autoref{sec:analytical_tasks}: 

\myparagraph{U1: Identify dominant attention elements} required toggling attention metrics to identify initial and sustained focus, mapping to \textbf{T1.1} and serving as a baseline for evaluating category-specific distributions (\textbf{T3.1}). 

\myparagraph{U2: Reconstruct visual trajectories} used the scarf plot and gaze overlays to describe sequential paths, operationalizing \textbf{T2.1} to validate the deconstruction of temporal scanpaths and transition episodes. 

\myparagraph{U3: Analyze specific urban features} involved inspecting regions of urban decay via glyphs and switching data (e.g., to Disorder Cues), addressing \textbf{T3.1} by linking localized fixations on physical disorder with safety judgments. 

\myparagraph{U4: Discover global viewing patterns} had users explore multiple images in the By-Participant view; this scenario covered the participant-centric tasks by evaluating cross-image attention (\textbf{T1.2}), temporal strategy shifts like fatigue (\textbf{T2.2}), and judgment consistency across visually similar embeddings (\textbf{T3.2}).

For each task, participants were encouraged to think aloud~\cite{lewis1982using}, verbalizing their reasoning and observations before answering quantitative (\textbf{QT}) and qualitative (\textbf{QL}) questions. Because the tasks were designed to support open-ended, autonomous discovery and hypothesis generation, we intentionally omitted conventional task-performance metrics, such as task completion rates and error frequencies. Such metrics do not adequately capture the qualitative synthesis and insight validation that are central to value-driven evaluation frameworks. The quantitative questions operationalized the ICE-T-inspired dimensions through eight statements (two per dimension), each rated on a 7-point Likert scale. The qualitative questions elicited open-ended feedback on the usefulness of the visual components and suggestions for improvement. The full questionnaire is provided in~\ref{appx:icet}.

\figIceChart

\subsection{Results}

Both quantitative results from our ICE-T-inspired questionnaire and qualitative feedback are reported below.


\subsubsection{Quantitative Results}

Averaged over the 22 participants, all four ICE-T dimensions scored above the neutral midpoint (4) on the 7-point Likert scale, with \textbf{Confidence} highest (6.3), followed by \textbf{Essence} (6.2), \textbf{Time} (6.1), and \textbf{Insight} (5.9); individual results are reported in~\ref{appx:icet_results}. We qualify these ratings by noting that our cohort consisted of general data analysts with no prior domain exposure to eye-tracking data; the high Confidence and Essence scores thus reflect the system's ability to make complex, aggregated spatiotemporal gaze data immediately accessible and trustworthy for non-expert users, rather than an expert-level verification of visualization heuristics.

\textbf{Insight (QT1, QT2)} showed more varied responses, suggesting deeper interpretation may rely on extended user familiarity. \textbf{Confidence (QT3, QT4)} showed strong agreement, as coordinated views and consistent color encodings enhanced trust in spatial and temporal interpretations. \textbf{Essence (QT5, QT6)} results highlighted the system's ability to summarize complex visual behavior concisely. The \textbf{Time (QT7, QT8)} dimension was rated positively, though temporal visualizations required slightly more cognitive effort. \autoref{fig:icet_chart} confirms most responses fall on the positive end of the Likert scale, with minimal neutral or negative feedback.

\subsubsection{Qualitative feedback}
Participant feedback (\textbf{QL1}, \textbf{QL2}) consistently highlighted the \textit{Participant Attention Summary} and the \textit{Temporal Participant Attention} scarf plot as the most intuitive components for comparing focus, interpreting numerical metrics, and tracking chronological scanpaths. Interactive features like cross-view filtering were deemed essential for reducing visual clutter; however, users noted that while the \textit{By-Image} mode was visually accessible, the \textit{By-Participant} mode required higher cognitive effort due to its numerical density. Improvement requests centered on expanding global search (e.g., quickly locating specific images or participants) and sorting results by attention-based metrics. Users also suggested minor refinements such as resizable panels, and proposed automated insights to highlight dominant attention correlations across scenes for higher-level analytical reasoning.
\section{Discussion}
\label{section:discussion} \label{sec:results}

The quantitative validation established in \autoref{section:dataset_validation} confirms that the \dataname dataset possesses a stable signal-to-noise ratio. In this section, we interpret the domain-specific implications of these patterns, discussing how visual attention aligns with foundational urban sociology and how fine-grained granularity reveals context-specific nuances.

\subsection{Alignment with Broken Windows Theory}

Consistent with the Broken Windows Theory~\cite{wilson1982Broken}, our hybrid regression metrics, detailed in \autoref{tab:lmm_disorder}, demonstrate that physical disorder cues from UrbanPD4k are the strongest negative predictors of perceived urban safety. Specifically, sustained visual attention to broken or damaged brick walls ($\beta = -0.442$, $p < .001$) demonstrated the most pronounced negative association with safety scores, closely followed by damaged pavement ($\beta = -0.229$) and graffiti ($\beta = -0.225$). These results confirm that the interactive visual patterns highlighted by analysts within the \systemname \textbf{Attention View} correspond directly to statistically meaningful and predictable human perceptual mechanisms of decay.

\begin{table}[htbp]
\centering
\caption{Fixed effects of visual attention on perceived safety (ADE20K + Disorder, $N = 1500$).}
\label{tab:lmm_disorder}
\footnotesize
\setlength{\tabcolsep}{3pt}
\begin{tabular}{l c c r c c c}
\toprule
\textbf{Predictor} & \textbf{Coef. ($\beta$)} & \textbf{S.E.} & \textbf{$z$-val} & \textbf{VIF} & \textbf{$p$-val} & \textbf{95\% CI} \\
\midrule
(Intercept) & 5.091 & 0.159 & 31.940 & -- & <.001 & [4.780, 5.400] \\
\midrule
\textit{Positive (Safe)} & & & & & & \\
Car / Taxi & 0.197 & 0.047 & 4.230 & 1.160 & <.001 & [0.110, 0.290] \\
Door & 0.155 & 0.051 & 3.060 & 1.360 & 0.002 & [0.060, 0.250] \\
\midrule
\textit{Negative (Unsafe)} & & & & & & \\
Broken/Dam Wall & -0.442 & 0.090 & -4.930 & 4.270 & <.001 & [-0.620, -0.270] \\
Wall & -0.326 & 0.121 & -2.690 & 7.780 & 0.007 & [-0.560, -0.090] \\
Road & -0.277 & 0.117 & -2.360 & 7.340 & 0.018 & [-0.510, -0.050] \\
Damaged Pavement & -0.229 & 0.056 & -4.100 & 1.650 & <.001 & [-0.340, -0.120] \\
Graffiti & -0.225 & 0.075 & -2.990 & 2.990 & 0.003 & [-0.370, -0.080] \\
Sky & -0.200 & 0.071 & -2.830 & 2.170 & 0.005 & [-0.340, -0.060] \\
Bridge & -0.188 & 0.051 & -3.690 & 1.370 & <.001 & [-0.290, -0.090] \\
Mountain & -0.184 & 0.046 & -4.020 & 1.120 & <.001 & [-0.270, -0.090] \\
Signboard & -0.152 & 0.055 & -2.770 & 1.590 & 0.006 & [-0.260, -0.040] \\
Garbage Bag & -0.142 & 0.049 & -2.900 & 1.280 & 0.004 & [-0.240, -0.050] \\
Pole & -0.142 & 0.048 & -2.990 & 1.210 & 0.003 & [-0.240, -0.050] \\
Field & -0.138 & 0.045 & -3.070 & 1.070 & 0.002 & [-0.230, -0.050] \\
\midrule
\textit{Random Effects} & \textbf{Var.} & \textbf{SD} & & \textbf{ICC} & & \\
Participant & 0.707 & 0.841 & & 0.203 & & \\
Residual & 2.774 & 1.666 & & & & \\
\bottomrule
\end{tabular}
\end{table}

\subsection{General Urban Infrastructure vs. Semantic Granularity}
Similar results for the other three data cases (detailed in the Supplementary Material~\ref{appx:detailed_lmm}) underscore the importance of semantic granularity. Baseline models (such as raw ADE20K) merge deterioration into broad, generic categories like \textit{wall}, whereas our ADE20K+UrbanPD4k separates general infrastructure from decay signals: the baseline absorbs structural decay into background elements, while our fine-grained representation distinguishes blind walls ($\beta = -0.326$) from physically damaged walls ($\beta = -0.442$), the latter showing a substantially stronger negative correlation.

Conversely, macro-level Grouped-ADE proved too coarse and statistically unstable for safety analysis. Variance Inflation Factor (VIF) analysis revealed severe multicollinearity when fine-grained elements were merged into broad clusters---for example, aggregating infrastructure into \textit{Construction} or ground elements into \textit{Floor} produced VIF values of 43.29 and 29.78, respectively, inflating standard errors enough to render several predictors insignificant. These findings support the necessity of fine-grained, disorder-aware semantic cues in urban visual analytics.

\subsection{The Mitigating Role of Social Cues}
Beyond physical decay, \systemname captures localized contextual patterns that global statistical models often obscure. As shown in our second usage scenario in \autoref{sec:scenario_2}, the visual presence of pedestrians acts as a powerful social mediator: even amid prominent disorder cues like graffiti, scenes with active human presence (Image ID 72) maintained substantially higher safety ratings than visually identical, isolated spaces (Image ID 10). This suggests that visual attention to human presence is consistent with natural surveillance interpretations, modifying how physical decay is interpreted. Across these qualitative cases, the presence of people often exerts a stronger positive effect on perceived safety than visual clutter or graffiti alone.
\section{Limitations and Future Work}

Although this study provides interpretable insights into urban safety perception, several limitations affect the scope and generalizability of the findings. First, regarding \textbf{participant diversity and experimental context}, our sample of 30 participants from a single institution is sufficient for methodological evaluation but does not represent the socioeconomic diversity of Rio de Janeiro residents. Moreover, the analysis relies on static SVI viewed for 15 seconds in a controlled head-mounted setting, whereas real-world perception is shaped by movement, sound, and unconstrained exploration. \textbf{Future work} should therefore include more diverse populations and multimodal, in-the-wild data collection.

Second, \systemname{} has limitations in \textbf{visual scalability, representation, and the correlational nature of attention}. Designed for a moderate dataset (150 images, 30 participants), visualizations such as heatmaps and scarf plots may require filtering or aggregation for larger cohorts. Similarly, Places365 embeddings capture coarse scene semantics but may miss fine-grained disorder cues, while t-SNE projections remain sensitive to initialization with only 50 images per participant. \textbf{Future work} should investigate scalable visualization and dimensionality-reduction strategies, as well as domain-specific visual representations that better capture fine-grained urban disorder cues. Finally, although attention to physical disorder is associated with lower safety ratings, these results are correlational rather than causal. \systemname{} is therefore intended as an exploratory, hypothesis-generating tool, motivating future neurocognitive studies with manipulated stimuli.

Third, regarding \textbf{spatial mapping}, \systemname{} does not include a macro-geographic map layer. Because the dataset prioritizes structural diversity over geographic coverage, mapping the sampled locations could imply unsupported spatial trends. \textbf{Future work} should integrate micro-level gaze analysis with city-scale geospatial representations to better connect perception and urban planning.

\section{Conclusions}

We presented \systemname, a visual analytics system for exploring eye-movement behavior in urban safety perception. Built on a head-mounted eye-tracking study with 30 participants and 150 SVIs from Rio de Janeiro, the system integrates gaze dynamics, semantic segments, disorder annotations, and safety ratings into a unified analytical framework.

Our results show that perceived safety cannot be fully understood through image-level judgments alone. By linking gaze behavior, temporal attention patterns, and inspected urban elements, \systemname{} reveals how disorder cues, infrastructure, vegetation, and human presence contribute differently to safety perception, complementing statistical analyses by exposing local, context-specific patterns often hidden by global aggregation.
More broadly, this work demonstrates the value of integrating eye tracking, semantic scene understanding, and visual analytics to support more interpretable studies of urban perception, providing a foundation for future research in urban analytics, human-centered planning, and perception-aware urban design.

\section*{ACKNOWLEDGMENTS}

This work was supported by the National Council for Scientific and Technological Development (CNPq, grant \#313454/2026-4 and \#132349/2025-6), Brazilian Federal Agency for Support and Evaluation of Graduate Education (CAPES, grant \#88887.684234/2022-00), Carlos Chagas Filho Foundation for Research Support of Rio de Janeiro State (FAPERJ, grant \#E-26/210.585/2025), São Paulo Research Foundation (FAPESP, grant \#2021/07012-0 and \#2024/05760-8),  and the Fundação Getulio Vargas (FGV).

\bibliographystyle{cag-num-names}
\bibliography{refs}

\clearpage
\newpage
\appendix

\setcounter{page}{1}
\setcounter{figure}{0}
\setcounter{table}{0}

\section*{Supplementary Material}

Supplementary material that may be helpful in the review process. 

\section{Multivariate Stratified Sampling Optimization} \label{appendix:math_form}
To preserve the diversity of visual element proportions, we used a multivariate stratified sampling scheme~\cite{khowaja2011estimation}. For each visual element $v \in V$, we compute its pixel-ratio distribution over all images and divide it into quartile groups $g \in G_v$. Let $S_{v,g}$ be the set of images whose proportion of element $v$ falls into quartile $g$, and let $s$ be the desired sample size. We select a binary vector $\mathbf{x} = (x_1,\dots,x_n)$, where $x_i = 1$ if image $i$ is kept, by solving:

\begin{gather}
    \text{objective:} \quad \max \sum_{i=1}^{n} x_i \label{eq:objective} \\
    \text{subject to:} \quad 
    \Bigl\lfloor \frac{s}{\sum_{v\in V}\lvert G_v\rvert}\Bigr\rfloor 
    \;\le\; \sum_{i\in S_{v,g}} x_i
    \quad \forall v \in V, \forall g \in G_v, \label{eq:constraints}
\end{gather}

which enforces a minimum number of selected images in each quartile across all visual elements, yielding the final 150-image subset.

\section{Randomized Image Assignment Matrix} 
\label{appendix:random_matrix}
Given $n_p$ participants $\mathcal{P}$ and $n_i$ selected images $\mathcal{I}$, we require that every image be rated by exactly $K_{\mathcal{P}}$ distinct participants and every participant view exactly $K_{\mathcal{I}}$ images, with $n_i K_{\mathcal{P}} = n_p K_{\mathcal{I}}$. In our study, we targeted $K_{\mathcal{P}} = 10$ ratings per image. We implement a randomized assignment procedure that iteratively fills an $n_p \times n_i$ binary matrix $V$ such that:

\begin{gather}
    \sum_{p=1}^{n_p} V_{p,i} = K_{\mathcal{P}} \quad \forall i \in \mathcal{I}, \label{eq:participant_constraint}\\
    \sum_{i=1}^{n_i} V_{p,i} = K_{\mathcal{I}} \quad \forall p \in \mathcal{P}, \label{eq:image_constraint}
\end{gather}

where $V_{p,i} = 1$ denotes that participant $p$ views image $i$.

\section{Extended Experimental Protocol and Hardware Setup} \label{appx:experimental_protocol}

\textbf{This study was conducted with approval from the Institutional Ethics Committee of our university. }
The experiment followed a standardized protocol with three main phases, lasting at most 25 minutes to avoid fatigue~\cite{tatler2011eye,clay2019eye}. \autoref{fig:perception_assessment} shows the complete experimental protocol for eye-gaze data collection.

\myparagraph{Calibration:} The HoloLens 2 device was cleaned between sessions. We performed a nine-point gaze calibration routine, monitoring accuracy in real time. Following established benchmarks, our setup maintained a quantitative spatial accuracy of $0.77^\circ \pm 0.35^\circ$ at the 2.0m focal distance used for stimuli projection.

\myparagraph{Task familiarization:} Participants viewed two reference images—one clearly perceived as \emph{unsafe}, one as \emph{safe}—for 15\,s each to confirm stable tracking.

\myparagraph{Safety rating trials:} To prevent visual carry-over effects and ensure stable semantic scenes, we implemented a 5-second interval buffer between stimuli. Any trial failing to meet the 15-second viewing duration due to system interruption was automatically discarded.

\figPerceptionAssessment

\section{ICE-T Questionnaire and Open-Ended Questions} \label{appx:icet}

The quantitative questions were directly \textit{inspired by} the ICE-T framework~\cite{wall2018heuristic}, covering four dimensions: Insight, Confidence, Essence, and Time. Each dimension was assessed through two statements using a 7-point Likert scale (from ``Totally Disagree'' to ``Totally Agree'').

\textbf{Insight.} 

\myparagraph{QT1}: ``\textit{In the By-Image view, the ability to switch between attention metrics and segmentation classes helped me identify relevant elements and generate new interpretations of visual attention.}''.

\myparagraph{QT2}: ``\textit{The combination of visual representations in the By-Image view with the metrics in the By-Participant view allowed me to discover patterns and better understand participants' exploration behavior.}''..

\textbf{Confidence.} 

\myparagraph{QT3}: ``\textit{In the By-Image view, the consistency of colors associated with semantic classes across visualizations, together with tooltips for inspecting exact values, increased my confidence in interpreting the data accurately.}''.

\myparagraph{QT4}: ``\textit{The synchronized coordination between visualizations and the overlay of gaze/fixation points on the image increased my confidence in the correctness of spatial and temporal alignment.}''.

\textbf{Essence.} 

\myparagraph{QT5}: ``\textit{Both image-centered analysis and participant-centered analysis facilitated a quick and overall understanding of visual behavior.}''.

\myparagraph{QT6}: ``\textit{The system effectively summarizes complex eye-tracking data by integrating multiple coordinated views, while maintaining clarity.}''.

\textbf{Time.} 

\myparagraph{QT7}: ``\textit{The use of temporal visualization tools allowed me to quickly relate when participants observed specific elements and identify dominant attention patterns over time.}''.

\myparagraph{QT8}: ``\textit{Filtering options and smooth navigation significantly reduced the effort and time required to derive insights from eye-tracking data.}''.

The qualitative questions collected open-ended feedback about the usefulness of the system and potential improvements:

``\textit{Considering both the By-Image and By-Participant views, which visual component, chart, or interaction did you find most useful for your analysis, and why?}'' (QL1);``\textit{What would you change, add, or remove from the interface to make it more efficient? Please feel free to include any additional comments or suggestions about your experience.}'' (QL2).

\section{ICE-T Results Per Participants} \label{appx:icet_results}

\autoref{tab:ice_table_participants} presents the ICE-T scores per participant, indicating consistently positive evaluations across all four dimensions.


\begin{table}[t!]
\centering
\caption{ICE-T evaluation scores per participant across the four dimensions.}
\label{tab:ice_table_participants}
\begin{tabular}{lcccc}
\toprule
Participant & \textit{\textbf{Insight}} & \textit{\textbf{Confidence}} & \textit{\textbf{Essence}} & \textit{\textbf{Time}} \\
\midrule
P1  & 6.5 & 7.0 & 7.0 & 7.0 \\
P2  & 6.0 & 6.0 & 6.0 & 6.0 \\
P3  & 5.5 & 7.0 & 6.0 & 5.5 \\
P4  & 3.5 & 4.0 & 6.0 & 5.0 \\
P5  & 5.0 & 6.5 & 5.0 & 6.5 \\
P6  & 6.5 & 7.0 & 6.0 & 6.5 \\
P7  & 5.0 & 6.5 & 6.5 & 7.0 \\
P8  & 5.5 & 5.5 & 5.5 & 6.5 \\
P9  & 6.0 & 6.0 & 6.0 & 6.0 \\
P10 & 5.5 & 7.0 & 6.5 & 6.5 \\
P11 & 6.0 & 6.0 & 5.5 & 5.0 \\
P12 & 5.5 & 6.5 & 6.0 & 4.5 \\
P13 & 6.5 & 6.5 & 6.0 & 7.0 \\
P14 & 7.0 & 7.0 & 6.5 & 7.0 \\
P15 & 6.0 & 5.5 & 6.0 & 6.5 \\
P16 & 5.5 & 5.5 & 6.0 & 5.5 \\
P17 & 7.0 & 6.5 & 6.0 & 6.0 \\
P18 & 7.0 & 7.0 & 7.0 & 5.0 \\
P19 & 7.0 & 6.0 & 7.0 & 7.0 \\
P20 & 6.5 & 7.0 & 7.0 & 7.0 \\
P21 & 4.5 & 7.0 & 7.0 & 7.0 \\
P22 & 6.0 & 5.0 & 6.5 & 4.0 \\
\midrule
\textbf{Avg.} 
& \textbf{5.9}
& \textbf{6.3}
& \textbf{6.2}
& \textbf{6.1} \\
\bottomrule
\end{tabular}
\vspace{-15pt}
\end{table}

\newpage

\section{Spatial Uncertainty and Kernel Calculation} \label{appx:Kernel_Calculation}

To ensure the statistical validity of the \textit{Attention Intensity} metric, it is necessary to calibrate the gaze data to account for the inherent hardware noise of the HoloLens 2. This appendix details the step-by-step derivation of the spatial kernel ($\sigma$) and the area exclusion threshold ($\tau$) used to prevent false-positive gaze allocations on minute urban elements (e.g., distant overhead cables or small debris).

\textbf{Step 1: Angular Precision of the Device} \\
In eye-tracking methodology, spatial \textit{precision} is the standard parameter used to define the dispersion of a Gaussian kernel, as it models the physiological and hardware jitter of a fixation. According to benchmark evaluations~\cite{kapp2021arett}, the HoloLens 2 exhibits a spatial precision of $0.24^\circ$ of visual angle at a focal distance of 2.0 meters, which is the exact distance used in our experimental setup.

\textbf{Step 2: Conversion to Pixels Per Degree (PPD)} \\
To translate this angular precision into the 2D pixel space of our stimuli, we calculated the device's angular resolution. Based on the manufacturer’s specifications\footnote{https://learn.microsoft.com/en-us/hololens/hololens2-hardware} and hardware evaluations, the HoloLens 2 displays a resolution of 2,048 $\times$ 1,080 pixels per eye with a diagonal field of view (FOV) of $52^\circ$ and a diagonal of 2,315 pixels. Dividing this by the diagonal FOV yields an angular resolution of approximately 44.5 Pixels Per Degree ($2,315 \text{ px} / 52^\circ \approx 44.5 \text{ PPD}$).

\textbf{Step 3: Deriving the Gaussian Kernel ($\sigma$)} \\
Multiplying the device's angular precision by the calculated PPD directly provides the operational standard deviation ($\sigma$) for our heatmap generation. Thus, we applied a Gaussian kernel with $\sigma \approx 10.68$ pixels ($0.24^\circ \times 44.5 \text{ PPD}$). This ensures that the spatial spread of each fixation accurately reflects the empirical uncertainty of the device.

\textbf{Step 4: Defining the Area Exclusion Threshold ($\tau$)} \\
Finally, to establish a robust noise threshold ($\tau$) for semantic segmentation, we considered an effective noise radius of $2\sigma$ ($2 \times 10.68 = 21.36$ pixels). In a 2D Gaussian distribution, a $2\sigma$ radius encompasses over 95.4\% of the spatial uncertainty. Geometrically, this margin of error forms a circular area of roughly 1,433 squared pixels ($A = \pi \times 21.36^2$).

When evaluated against our $800 \times 600$ stimulus resolution (480,000 total pixels), this hardware noise floor occupies exactly $0.3\%$ of the visual field ($1,433 / 480,000 \approx 0.003$). To ensure an even more robust and conservative analysis, we padded this noise floor with a safety margin, establishing a strict area threshold of $\tau = 0.005$ ($0.5\%$). Any semantic mask or disorder cue occupying less than this threshold was considered statistically indistinguishable from hardware noise and was excluded from the analysis.

\section{Inter-Rater Reliability (IRR) and Rater Calibration} \label{appx:Reliability}

To validate the consistency of urban safety perceptions and ensure that our findings reflect environmental features rather than individual biases, we performed a reliability analysis tailored for sparse rating matrices. Given our \textit{Ill-Structured Measurement Design} (ISMD)—where 30 participants evaluated overlapping but incomplete subsets of 150 images—traditional Intraclass Correlation Coefficients (ICC) based on balanced ANOVA are mathematically biased. 

Following the framework proposed by Putka et al. (2008), we utilized a Linear Mixed-Effects Model (LMM) with crossed random effects to isolate the variance components of our dataset ($N=1,500$ observations). This approach allows for explicit rater calibration by partitioning the total variance into three distinct sources: the urban scene (signal), the participant (subjective bias), and residual noise. The variance components estimated via Restricted Maximum Likelihood (REML) are detailed in \autoref{tab:variance_components}.

\begin{table}[h]
\centering
\caption{Variance Components for Safety Perception Scores.}
\label{tab:variance_components}
\resizebox{\columnwidth}{!}{%
\begin{tabular}{lcl}
\hline
\textbf{Source of Variation} & \textbf{Variance ($\sigma^2$)} & \textbf{Interpretation} \\ \hline
Urban Scene (Image) & 0.876 & True variance in streetscape quality (Signal) \\
Participant (Rater) & 0.394 & Subjective severity/leniency bias (Calibration) \\
Residual (Noise) & 1.766 & Unexplained variance and interaction \\ \hline
\end{tabular}%
}
\end{table}

To calculate the reliability of the aggregated safety scores, we applied the $G(q,k)$ estimator, which accounts for the partial overlap between raters in ISMD designs:

\begin{equation}
G(q,k) = \frac{\sigma^2_{img}}{\sigma^2_{img} + \frac{q \cdot \sigma^2_{part} + \sigma^2_{res}}{k}}
\end{equation}

Where $k=10$ represents the number of raters per image, and $q \approx 0.67$ is the coefficient of non-overlap for our design. This coefficient is mathematically derived from the ratio of raters per image to the total participant pool ($q = 1 - k/N_{total}$), reflecting that each image was evaluated by exactly one-third of our 30-participant panel ($1 - 10/30$). 

The analysis yielded a high inter-rater reliability of \textbf{$G(q,k) = 0.812$}. In the context of behavioral research, a value exceeding 0.80 indicates a robust and high level of consensus. This result confirms that the safety metrics used in \textsc{UrbanGazeVis} are highly stable and that the sample of 10 raters per image provides sufficient statistical power to capture the collective perception of the urban environment.

\section{Detailed Statistical Results for LMM} \label{appx:detailed_lmm}
\setcounter{table}{0}

This section provides the comprehensive statistical outputs for the four semantic data evaluated in this study: \textit{ADE20K Base}, \textit{Grouped-ADE}, \textit{ADE20K + UrbanPD4k}, and \textit{Grouped-ADE + UrbanPD4k}. While the main text focuses on the most predictive features, the following tables provide the full set of fixed effects, including non-significant predictors, to ensure transparency and allow for meta-analytical comparisons. 

To focus the analysis on strictly structural and urban elements, broad contextual categories (such as indoor, outdoor, nature, and miscellaneous) have been excluded from these summary tables. Each table reports the estimated coefficients ($\beta$), standard errors (S.E.), $z$-values, Variance Inflation Factors (VIF, where applicable), $p$-values, and 95\% confidence intervals (CI) for the visual elements. Predictors are categorized into elements that positively correlate with perceived safety (Safe) and those that correlate negatively (Unsafe).

\subsection{ADE20K Base Model Results}
\autoref{tab:lmm_base_full} presents the results for the baseline ADE20K classes. This model serves as the primary benchmark for evaluating the impact of semantic granularity and disorder cues.

\begin{table*}[htbp]
\centering
\caption{Report: ADE20K (baseline)}
\label{tab:lmm_base_full}
\begin{threeparttable}
\begin{tabular}{lcccccc}
\toprule
\textbf{Predictor} & \textbf{Coef. ($\beta$)} & \textbf{S.E.} & \textbf{z-val} & \textbf{VIF} & \textbf{p-val} & \textbf{95\% CI} \\
\midrule
(Intercept) & 5.091$^{***}$ & 0.155 & 32.790 & --- & $< .001$ & [4.790,\ 5.400] \\
\midrule
\multicolumn{7}{l}{\textit{Positive (Safe)}} \\
\quad Tree        &  0.195          & 0.084 &  2.310 &  3.400 & 0.056 & [0.030,\ 0.360]   \\
\quad Door        &  0.182$^{**}$   & 0.051 &  3.570 &  1.300 & 0.010 & [0.080,\ 0.280]   \\
\quad Palm        &  0.130          & 0.056 &  2.320 &  1.560 & 0.056 & [0.020,\ 0.240]   \\
\quad Ashcan      &  0.113          & 0.045 &  2.500 &  1.020 & 0.055 & [0.020,\ 0.200]   \\
\quad Fence       &  0.099          & 0.093 &  1.060 &  4.290 & 0.398 & [-0.080,\ 0.280]  \\
\quad Plant       &  0.097          & 0.055 &  1.760 &  1.520 & 0.162 & [-0.010,\ 0.200]  \\
\quad Bus         &  0.096          & 0.050 &  1.910 &  1.250 & 0.135 & [-0.000,\ 0.190]  \\
\quad Minibike    &  0.078          & 0.047 &  1.650 &  1.130 & 0.190 & [-0.010,\ 0.170]  \\
\quad Streetlight &  0.063          & 0.046 &  1.390 &  1.040 & 0.262 & [-0.030,\ 0.150]  \\
\quad Car         &  0.036          & 0.076 &  0.470 &  2.910 & 0.719 & [-0.110,\ 0.190]  \\
\quad Sidewalk    &  0.004          & 0.098 &  0.040 &  4.790 & 0.970 & [-0.190,\ 0.200]  \\
\midrule
\multicolumn{7}{l}{\textit{Negative (Unsafe)}} \\
\quad House       & -0.003          & 0.054 & -0.060 &  1.470 & 0.970 & [-0.110,\ 0.100]  \\
\quad Van         & -0.016          & 0.048 & -0.330 &  1.170 & 0.803 & [-0.110,\ 0.080]  \\
\quad Person      & -0.030          & 0.055 & -0.550 &  1.500 & 0.687 & [-0.140,\ 0.080]  \\
\quad Truck       & -0.062          & 0.066 & -0.940 &  2.180 & 0.448 & [-0.190,\ 0.070]  \\
\quad Stairs      & -0.064          & 0.046 & -1.390 &  1.050 & 0.262 & [-0.150,\ 0.030]  \\
\quad Grass       & -0.072          & 0.060 & -1.190 &  1.820 & 0.352 & [-0.190,\ 0.050]  \\
\quad Earth       & -0.102          & 0.097 & -1.050 &  4.740 & 0.398 & [-0.290,\ 0.090]  \\
\quad Signboard   & -0.105          & 0.056 & -1.880 &  1.560 & 0.135 & [-0.220,\ 0.000]  \\
\quad Field       & -0.130$^{*}$    & 0.046 & -2.800 &  1.070 & 0.027 & [-0.220,\ -0.040] \\
\quad Mountain    & -0.153$^{*}$    & 0.047 & -3.260 &  1.110 & 0.013 & [-0.250,\ -0.060] \\
\quad Pole        & -0.154$^{*}$    & 0.049 & -3.180 &  1.170 & 0.013 & [-0.250,\ -0.060] \\
\quad Bridge      & -0.159$^{*}$    & 0.052 & -3.060 &  1.360 & 0.015 & [-0.260,\ -0.060] \\
\quad Building    & -0.179          & 0.219 & -0.820 & 24.020 & 0.509 & [-0.610,\ 0.250]  \\
\quad Sky         & -0.188          & 0.077 & -2.430 &  2.450 & 0.055 & [-0.340,\ -0.040] \\
\quad Road        & -0.198          & 0.122 & -1.620 &  7.520 & 0.190 & [-0.440,\ 0.040]  \\
\quad Wall        & -0.370          & 0.154 & -2.400 & 11.820 & 0.055 & [-0.670,\ -0.070] \\
\midrule
\multicolumn{7}{l}{\textit{Random Effects}} \\
& \textbf{Var.} & \textbf{SD} & \multicolumn{2}{c}{\textbf{ICC}} & & \\
\quad Participant & 0.665 & 0.815 & \multicolumn{2}{c}{\multirow{2}{*}{0.184}} & & \\
\quad Residual    & 2.941 & 1.715 & & & & \\
\bottomrule
\end{tabular}
\begin{tablenotes}
\small
\item $^{*}\ p < 0.05$, $^{**}\ p < 0.01$, $^{***}\ p < 0.001$
\end{tablenotes}
\end{threeparttable}
\end{table*}

\subsection{Grouped-ADE Model Results}
\autoref{tab:lmm_grouped_full} details the results for the macro-level Grouped-ADE where elements are clustered into broad infrastructure categories. As discussed in \autoref{sec:results}, this model exhibits significant multicollinearity in certain categories.

\begin{table*}[htbp]
\centering
\caption{Report: Grouped-ADE}
\label{tab:lmm_grouped_full}
\begin{threeparttable}
\begin{tabular}{lcccccc}
\toprule
\textbf{Predictor} & \textbf{Coef. ($\beta$)} & \textbf{S.E.} & \textbf{z-val} & \textbf{VIF} & \textbf{p-val} & \textbf{95\% CI} \\
\midrule
(Intercept) & 5.091$^{***}$ & 0.159 & 32.080 & --- & $< .001$ & [4.780,\ 5.400] \\
\midrule
\multicolumn{7}{l}{\textit{Positive (Safe)}} \\
\quad Vegetation      &  0.262 & 0.149 &  1.760 & 10.210 & 0.288 & [-0.030,\ 0.550] \\
\quad Terrain Vehicle &  0.097 & 0.136 &  0.710 &  8.550 & 0.781 & [-0.170,\ 0.360] \\
\quad Human           &  0.029 & 0.065 &  0.450 &  1.930 & 0.851 & [-0.100,\ 0.160] \\
\midrule
\multicolumn{7}{l}{\textit{Negative (Unsafe)}} \\
\quad Construction  & -0.057 & 0.306 & -0.190 & 43.290 & 0.851 & [-0.660,\ 0.540]  \\
\quad Floor         & -0.073 & 0.254 & -0.290 & 29.780 & 0.851 & [-0.570,\ 0.420]  \\
\quad City Elements & -0.107 & 0.070 & -1.530 &  2.260 & 0.346 & [-0.240,\ 0.030]  \\
\quad Sky           & -0.178 & 0.099 & -1.790 &  4.110 & 0.288 & [-0.370,\ 0.020]  \\
\midrule
\multicolumn{7}{l}{\textit{Random Effects}} \\
& \textbf{Var.} & \textbf{SD} & \multicolumn{2}{c}{\textbf{ICC}} & & \\
\quad Participant & 0.692 & 0.832 & \multicolumn{2}{c}{\multirow{2}{*}{0.179}} & & \\
\quad Residual    & 3.180 & 1.783 & & & & \\
\bottomrule
\end{tabular}
\begin{tablenotes}
\small
\item $^{*}\ p < 0.05$, $^{**}\ p < 0.01$, $^{***}\ p < 0.001$
\end{tablenotes}
\end{threeparttable}
\end{table*}

\subsection{ADE20K + UrbanPD4k Model Results}
\autoref{tab:lmm_disorder_full} provides the complete output for the hybrid model integrating specific physical disorder cues. This model incorporates predictors such as ``Broken/Damaged Wall'' and ``Graffiti'', which emerged as critical indicators of perceived insecurity.

\begin{table*}[htbp]
\centering
\caption{Report: ADE20K + UrbanPD4k}
\label{tab:lmm_disorder_full}
\begin{threeparttable}
\begin{tabular}{lcccccc}
\toprule
\textbf{Predictor} & \textbf{Coef. ($\beta$)} & \textbf{S.E.} & \textbf{z-val} & \textbf{VIF} & \textbf{p-val} & \textbf{95\% CI} \\
\midrule
(Intercept) & 5.091$^{***}$ & 0.159 & 31.940 & --- & $< .001$ & [4.780,\ 5.400] \\
\midrule
\multicolumn{7}{l}{\textit{Positive (Safe)}} \\
\quad Car / Taxi    &  0.197$^{***}$ & 0.047 &  4.230 &  1.160 & $< .001$ & [0.110,\ 0.290]   \\
\quad Door          &  0.155$^{*}$   & 0.051 &  3.060 &  1.360 & 0.010    & [0.060,\ 0.250]   \\
\quad Tree          &  0.100         & 0.083 &  1.200 &  3.510 & 0.302    & [-0.060,\ 0.260]  \\
\quad Plant         &  0.078         & 0.053 &  1.460 &  1.520 & 0.238    & [-0.030,\ 0.180]  \\
\quad Trashcan      &  0.076         & 0.044 &  1.720 &  1.040 & 0.164    & [-0.010,\ 0.160]  \\
\quad Palm          &  0.076         & 0.055 &  1.390 &  1.580 & 0.245    & [-0.030,\ 0.180]  \\
\quad Streetlight   &  0.052         & 0.045 &  1.170 &  1.050 & 0.308    & [-0.040,\ 0.140]  \\
\quad Motorcycle    &  0.041         & 0.046 &  0.890 &  1.140 & 0.412    & [-0.050,\ 0.130]  \\
\quad Bus           &  0.041         & 0.049 &  0.840 &  1.270 & 0.429    & [-0.060,\ 0.140]  \\
\quad Fence         &  0.013         & 0.089 &  0.150 &  4.180 & 0.884    & [-0.160,\ 0.190]  \\
\midrule
\multicolumn{7}{l}{\textit{Negative (Unsafe)}} \\
\quad Van              & -0.047          & 0.047 & -1.010 &  1.170 & 0.370    & [-0.140,\ 0.040]  \\
\quad House            & -0.050          & 0.053 & -0.940 &  1.500 & 0.395    & [-0.150,\ 0.050]  \\
\quad Sidewalk         & -0.061          & 0.094 & -0.650 &  4.600 & 0.532    & [-0.240,\ 0.120]  \\
\quad Stairs           & -0.068          & 0.045 & -1.520 &  1.060 & 0.224    & [-0.150,\ 0.020]  \\
\quad Car              & -0.083          & 0.074 & -1.130 &  2.870 & 0.315    & [-0.230,\ 0.060]  \\
\quad Overhead Cable   & -0.088          & 0.054 & -1.640 &  1.500 & 0.184    & [-0.190,\ 0.020]  \\
\quad Truck            & -0.089          & 0.065 & -1.370 &  2.250 & 0.245    & [-0.220,\ 0.040]  \\
\quad Person           & -0.095          & 0.054 & -1.770 &  1.530 & 0.159    & [-0.200,\ 0.010]  \\
\quad Grass            & -0.116          & 0.060 & -1.940 &  1.880 & 0.116    & [-0.230,\ 0.000]  \\
\quad Ground           & -0.117          & 0.090 & -1.300 &  4.330 & 0.268    & [-0.290,\ 0.060]  \\
\quad Field            & -0.138$^{*}$    & 0.045 & -3.070 &  1.070 & 0.010    & [-0.230,\ -0.050] \\
\quad Garbage Bag      & -0.142$^{*}$    & 0.049 & -2.900 &  1.280 & 0.012    & [-0.240,\ -0.050] \\
\quad Pole             & -0.142$^{*}$    & 0.048 & -2.990 &  1.210 & 0.010    & [-0.240,\ -0.050] \\
\quad Signboard        & -0.152$^{*}$    & 0.055 & -2.770 &  1.590 & 0.015    & [-0.260,\ -0.040] \\
\quad Mountain         & -0.184$^{***}$  & 0.046 & -4.020 &  1.120 & $< .001$ & [-0.270,\ -0.090] \\
\quad Bridge           & -0.188$^{**}$   & 0.051 & -3.690 &  1.370 & 0.001    & [-0.290,\ -0.090] \\
\quad Sky              & -0.200$^{*}$    & 0.071 & -2.830 &  2.170 & 0.014    & [-0.340,\ -0.060] \\
\quad Graffiti         & -0.225$^{*}$    & 0.075 & -2.990 &  2.990 & 0.010    & [-0.370,\ -0.080] \\
\quad Damaged Pavement & -0.229$^{***}$  & 0.056 & -4.100 &  1.650 & $< .001$ & [-0.340,\ -0.120] \\
\quad Road             & -0.277$^{*}$    & 0.117 & -2.360 &  7.340 & 0.043    & [-0.510,\ -0.050] \\
\quad Building         & -0.293          & 0.204 & -1.430 & 22.120 & 0.238    & [-0.690,\ 0.110]  \\
\quad Wall             & -0.326$^{*}$    & 0.121 & -2.690 &  7.780 & 0.018    & [-0.560,\ -0.090] \\
\quad Broken/Dam. Wall & -0.442$^{***}$  & 0.090 & -4.930 &  4.270 & $< .001$ & [-0.620,\ -0.270] \\
\midrule
\multicolumn{7}{l}{\textit{Random Effects}} \\
& \textbf{Var.} & \textbf{SD} & \multicolumn{2}{c}{\textbf{ICC}} & & \\
\quad Participant & 0.707 & 0.841 & \multicolumn{2}{c}{\multirow{2}{*}{0.203}} & & \\
\quad Residual    & 2.774 & 1.666 & & & & \\
\bottomrule
\end{tabular}
\begin{tablenotes}
\small
\item $^{*}\ p < 0.05$, $^{**}\ p < 0.01$, $^{***}\ p < 0.001$
\end{tablenotes}
\end{threeparttable}
\end{table*}

\subsection{Grouped-ADE + UrbanPD4k Model Results}
Finally, \autoref{tab:lmm_grouped_disorder_full} presents the results for the grouped classes augmented with disorder categories. This model evaluates whether the predictive power of disorder cues persists when environmental infrastructure is analyzed at a macro level.

\begin{table*}[htbp]
\centering
\caption{Report: Grouped-ADE + UrbanPD4k}
\label{tab:lmm_grouped_disorder_full}
\begin{threeparttable}
\begin{tabular}{lcccccc}
\toprule
\textbf{Predictor} & \textbf{Coef. ($\beta$)} & \textbf{S.E.} & \textbf{z-val} & \textbf{VIF} & \textbf{p-val} & \textbf{95\% CI} \\
\midrule
(Intercept) & 5.091$^{***}$ & 0.161 & 31.700 & --- & $< .001$ & [4.780,\ 5.410] \\
\midrule
\multicolumn{7}{l}{\textit{Positive (Safe)}} \\
\quad Car / Taxi      &  0.231$^{***}$ & 0.050 &  4.590 &  1.270 & $< .001$ & [0.130,\ 0.330]  \\
\quad Vegetation      &  0.156         & 0.138 &  1.130 &  9.500 & 0.425    & [-0.110,\ 0.430] \\
\quad Trashcan        &  0.085         & 0.046 &  1.850 &  1.050 & 0.128    & [-0.000,\ 0.170] \\
\midrule
\multicolumn{7}{l}{\textit{Negative (Unsafe)}} \\
\quad Terrain Vehicle    & -0.024          & 0.124 & -0.190 &  7.710 & 0.896    & [-0.270,\ 0.220]  \\
\quad Human              & -0.038          & 0.061 & -0.620 &  1.870 & 0.690    & [-0.160,\ 0.080]  \\
\quad Overhead Cable     & -0.082          & 0.059 & -1.380 &  1.740 & 0.299    & [-0.200,\ 0.030]  \\
\quad Construction       & -0.104          & 0.275 & -0.380 & 37.730 & 0.848    & [-0.640,\ 0.440]  \\
\quad City Elements      & -0.128          & 0.066 & -1.930 &  2.190 & 0.120    & [-0.260,\ 0.000]  \\
\quad Garbage Bag        & -0.142$^{*}$    & 0.054 & -2.630 &  1.460 & 0.031    & [-0.250,\ -0.040] \\
\quad Floor              & -0.158          & 0.226 & -0.700 & 25.460 & 0.690    & [-0.600,\ 0.290]  \\
\quad Sky                & -0.180          & 0.086 & -2.100 &  3.260 & 0.107    & [-0.350,\ -0.010] \\
\quad Graffiti           & -0.183          & 0.095 & -1.930 &  4.450 & 0.120    & [-0.370,\ 0.000]  \\
\quad Damaged Pavement   & -0.204$^{*}$    & 0.067 & -3.040 &  2.240 & 0.011    & [-0.330,\ -0.070] \\
\quad Broken/Dam. Wall   & -0.382$^{*}$    & 0.122 & -3.130 &  7.440 & 0.011    & [-0.620,\ -0.140] \\
\midrule
\multicolumn{7}{l}{\textit{Random Effects}} \\
& \textbf{Var.} & \textbf{SD} & \multicolumn{2}{c}{\textbf{ICC}} & & \\
\quad Participant & 0.715 & 0.846 & \multicolumn{2}{c}{\multirow{2}{*}{0.196}} & & \\
\quad Residual    & 2.943 & 1.716 & & & & \\
\bottomrule
\end{tabular}
\begin{tablenotes}
\small
\item $^{*}\ p < 0.05$, $^{**}\ p < 0.01$, $^{***}\ p < 0.001$
\end{tablenotes}
\end{threeparttable}
\end{table*}

\newpage
\clearpage

\section{Segmentation Model Comparison} \label{appendix:segmentation_model}

\autoref{fig:segmentation_model} shows the segmentation mask obtained from these three models. We note that SegFormer, PSPNet, and DeepLab do not segment the light pole; they fail to differentiate among walls, sidewalks, and gates, and they fail to segment cars correctly. However, while it requires more computational resources, the superior accuracy and versatility of OneFormer often justify the additional effort.

\figSegmentations


\end{document}